\documentstyle[12pt,dina4p]{article}
\renewcommand{\baselinestretch}{1.4}
\newcommand{\be}{\begin{equation}}
\newcommand{\ee}{\end{equation}}
\newcommand{\bear}{\begin{eqnarray}}
\newcommand{\ear}{\end{eqnarray}}

\newcommand{\Tr}{{\rm{Tr}}}

\newcommand{\D}{{\cal D}}
\newcommand{\A}{{\cal A}}

\date{}
\begin{document}
\begin{titlepage}
\begin{flushright}
DESY 96--081  \\
HD--THEP--96--11 \\
        hep-th/9605039
\end{flushright}
\quad\\
\vspace{1.8cm}
\begin{center}
{\LARGE \bf       Quantum Liouville Field Theory as}\\
\medskip
{\LARGE \bf          Solution of a Flow Equation}  \\
\vspace{1cm}
{\bf M. Reuter}\\
\bigskip
Deutsches Elektronen-Synchrotron DESY\\
Notkestrasse 85, D-22603 Hamburg \\
\vspace{.5cm}
{\bf C. Wetterich}\\
\bigskip
Institut  f\"ur Theoretische Physik\\
Universit\"at Heidelberg\\
Philosophenweg 16, D-69120 Heidelberg\\
\vspace{1cm}
{\bf Abstract}
\end{center}
A general framework for the Weyl invariant quantization
of Liouville field theory by means of an exact renormalization
group equation is proposed. This flow equation describes
the scale dependence of the effective average action which has a
built-in infrared cutoff.
For $c<1$~it is solved approximately by a truncation of the space
of action functionals. We derive the Ward identities associated to Weyl
transformations in presence of the infrared cutoff. They are used
to select a specific universality class for the renormalization
group trajectory which is found to connect two conformal
field theories with central charges $25-c$ and $26-c$, respectively.

\end{titlepage}
\newpage

\section{Introduction}

In spite of the fact that all its classical solutions are known
for more than a century \cite{li}, a satisfactory treatment of
quantum Liouville field theory is still lacking. Although a variety
of approaches, including canonical quantization
\cite{can}, \cite{nev}, \cite{ow}, \cite{nic},
inverse scattering techniques \cite{dj}, path integral \cite{gl},
quantum group \cite{ger}, \cite{js} and topological field
theory methods \cite{top}, have provided important insights into the
structure of the theory, it is still not possible to calculate
arbitrary $n$-point functions or to get a handle on the
notoriously difficult question about the structure of its
vacuum. Liouville theory is certainly a fascinating topic
also in its own right, but much of the recent interest in
this theory is due to its relevance to noncritical string
theory \cite{polya} and to 2-dimensional quantum gravity.
For $c>25$ it was also suggested to model essential
aspects of 4-dimensional quantum gravity \cite{polch}. The
subject was further stimulated by the comparison with
alternative approaches such as quantum gravity in the light-cone
gauge \cite{kpz}, \cite{cham} or matrix models \cite{mamo}.

In this paper we shall quantize Liouville field theory within the
general framework of the exact renormalization group approach
\cite{wil}. This method has led to an improved qualitative
understanding and conceptual clarification of many issues in
 renormalization theory.
It has been extended in various directions and its relation with
the concept of a coarse-grained free energy has been clarified
\cite{exrg}-\cite{ehw}. More precisely, we are
using a formulation in terms of the effective average action
\cite{avact}-\cite{cop}. The
average action $\Gamma_k$ \cite{avact} is a modification
of the standard effective action $\Gamma$ with a built-in
infrared cutoff $k$.\footnote{For a review see ref. \cite{cop}.}
It is the effective action appropriate for fields which have
been averaged over  spacetime volumes of size $k^{-1}$. Stated differently,
$\Gamma_k$ obtains from the classical action $S$ by integrating
out only the field modes with momenta larger than $k$, but
excluding the modes with momenta smaller than $k$. In the
limit $k\to0$, $\Gamma_k$ approaches the standard effective action,
$\Gamma_{k\to0}=\Gamma$. Here $\Gamma$ is the generating functional
for the 1PI Green's functions and includes all quantum fluctuations.
The dependence of $\Gamma_k$ on the scale $k$ is described by an exact
non-perturbative flow equation \cite{wet},\cite{gau}. This functional
differential equation takes the form of a
renormalization group-improved one-loop equation \cite{avact},
\cite{wet}, where the classical inverse propagator is replaced
by the full inverse propagator, i.e., the second functional
derivative of $\Gamma_k$ with respect to the fields.

The quantization of a theory in terms of the effective average action
proceeds in the following steps: One first specifies the
``short-distance action'' or ``microscopic action'' $S=\Gamma_\Lambda$
as the ``initial value'' of the average action specified at some
large scale $k=\Lambda$. This specification regularizes the
theory completely if the infrared cutoff is efficient enough, i.e.,
if the $k$-derivative of $\Gamma_k$ is ultraviolet finite.
One may consider the limit $\Lambda\to\infty$, but this is
not necessary and it is often more instructive to keep a finite
``ultraviolet cutoff scale'' $\Lambda$. On the other hand, the
quantum theory is completely given by the effective action
$\Gamma$. Quantization of a theory amounts then to the solution
of the flow equation for $k\to0$, starting with initial values
at the scale $\Lambda$. This procedure is rather general, but for
its practical implementation a few comments are in order.

i) While the flow equation is exact, it is usually not
possible to find exact solutions of it. There exists, however,
a rather attractive approximation scheme which is nonperturbative
in nature and does not require a small expansion parameter.
This method consists of truncating the space of all action
functionals, and to project the renormalization group flow
on a finite dimensional subspace. It is this technique of
projected renormalization group equations which we shall apply
to Liouville theory.

ii) In consequence, there are two basic ingredients in this
approach. The first is the existence of an exact flow equation
which can serve as a starting point for the approximations.
On this level our formulation is not much different from many
earlier versions of exact renormalization group equations
\cite{wil},\cite{exrg}, since all these formulations can be
related by suitable transformations.
The second ingredient is the possibility of an efficient
truncation. This means that the space of truncated actions must be
small enough to permit a computation and large enough to contain
all ``relevant physics''. For a judgement if the
``relevant physics'' is described by a given truncation, the
exact formulation of the effective action and the specific form of
the flow equation are of crucial importance. We believe that it is
useful in this respect to concentrate the emphasis on the variation
of an infrared cutoff since the problematic parts of the
``quantization'' typically involve a complicated infrared behavior.
This can be understood easily in the case of massless three-dimensional
scalar theories which are superrenormalizable in the ultraviolet,
but have a complicated infrared behavior
with fixed points and critical exponents. In addition, the most
important general properties of $\Gamma_k$ must be known
in order not to omit important effects by the truncation. For this
purpose, the free energy with an infrared cutoff, $\Gamma_k$, is a very
useful quantity since one often has some experience of the
general effects of quantum fluctuations and of the consequences
if the small momentum fluctuations $(q^2<k^2)$ are omitted. The
close analogy of our flow equation with renormalization group
improved one-loop perturbation theory helps considerably in this
respect.\footnote{Note that a relatively simple form of the
effective average action $\Gamma_k$ (which is 1PI-irreducible)
typically results in a quite complicated form of the
``ultraviolet cutoff action'' (1PI reducible) whose evolution is
described by earlier versions of exact renormalization group
equations \cite{wil}. In presence of massless particles and in
the momentum range $q^2\gg k^2$ a local form of $\Gamma_k$ corresponds
to a nonlocal ``ultraviolet cutoff action''. This makes it very
difficult to find non-perturbative truncations for
the ``ultraviolet cutoff action''. This problem is probably
the reason why most earlier attempts to solve the exact
renormalization group equations have remained in a perturbative
context (small coupling, small $\epsilon=4-d$, large number
$N$ of components). For more complicated theories such as non-abelian
gauge theories it is in addition notoriously difficult to
formulate a satisfactory ultraviolet cutoff.}

iii) The infrared cutoff must be quadratic in the quantum fields
$\chi$ as given by an additional piece in the action
\be\label{1.a}
\Delta_kS=\frac{1}{2}\int\frac{d^dq}{(2\pi)^d}\bar\chi
(q){\cal R}_k(q,q')\chi(q')\ee
with ${\cal R}_k$ typically given by ${\cal R}_k=R_k(q)(2\pi)^d
\delta(q-q')$ and $R_k(q)$ a function of $q^2$ that should
vanish identically for $k\to0$. This results in the flow
equation $(\partial_t=\partial/\partial\ln k)$
\be\label{1.b}
\partial_t\Gamma_k=\frac{1}{2}{\rm Tr}\{(\partial_t
{\cal R}_k)(\Gamma_k^{(2)}+{\cal R}_k)^{-1}\}\ee
with ${\rm Tr}$ containing a momentum integral. In four
dimensions a masslike infrared cutoff $R_k=Zk^2$ is usually
not sufficient to guarantee ultraviolet finiteness of the
r.h.s.. (More precisely, only the field-dependent part of $\Gamma_k$
is of interest.) Our proposal requires that
$R_k$ falls off sufficiently fast for large $q^2/k^2$,
and one may consider as an
example $\sim \exp(-q^2/k^2)$. In lower dimensions it is also
possible for some questions to use a masslike cutoff.

iv) It is not always possible to find a quadratic form
$\Delta_kS$ (\ref{1.a}) consistent with all symmetries of
the theory. In our case this will be relevant for the
Weyl-invariance of the Liouville theory. The effective
action $\Gamma_k$ will not exhibit all symmetries then. They are
recovered only for $k\to 0$ where the ``symmetry breaking term''
$\Delta_kS$ vanishes. In consequence, the Ward identity
for $\Gamma_k$ contains an additional
$k$-dependent  term which
vanishes only for $k=0$. It is very important that the
initial value $\Gamma_\Lambda$ is consistent with the anomalous
Ward identity\footnote{In a background field formalism
the Ward identity may re replaced by an exact identity for
the background dependence of $\Gamma_k[21]$.}
 - at least in a given approximation. Otherwise
the trajectory in the space of actions described by the solution
of the flow equation belongs to a different universality class
within which the required symmetries are not preserved even for $k=0$.

Before embarking on the  renormalization group
 equation and its solution,
let us briefly recall some basic facts about Liouville
theory which we shall need later on. We start from an arbitrary
conformal field theory of central charge $c$ which is coupled
to gravity and is governed by an action
$S_{\rm matter}[\Psi;g_{\mu\nu}]$. (In bosonic string theory the
matter fields $\Psi(x^\mu)$ correspond to the string positions
$X^m(x^\mu), m=1,...,c,\mu=1,2.$) Integrating out the matter field
fluctuations,
\be\label{1.1}
e^{-\Gamma_{\rm ind}[g]}=\int {\cal D}\Psi e^{-S_{\rm matter}
[\Psi;g]},\ee
leads to the induced gravity action \cite{polya}
\be\label{1.2}
\Gamma_{\rm ind}[g]=\frac{c}{96\pi}I[g]+\lambda\int d^2x\sqrt g\ee
with
\be\label{1.3}
I[g]=\int d^2x\sqrt gR\Delta_g^{-1}R\ee
where $\Delta_g$ is the Laplace-Beltrami operator of the metric
$g_{\mu\nu}$ and $R$  is the curvature scalar. In a second step one
has to integrate over the metric $g_{\mu\nu}$. We shall do
this in the conformal gauge by picking a reference metric $\hat g
_{\mu\nu}$ and writing
\be\label{1.4}
g_{\mu\nu}(x)=e^{2\phi(x)}\hat g_{\mu\nu}(x)\ee
We insert (\ref{1.4}) into (\ref{1.2}) and employ the
identity
\be\label{1.5}
I[e^{2\sigma}\hat g_{\mu\nu}]=I[\hat g_{\mu\nu}]-4\int d^2x
\sqrt{\hat g}\{\hat D_\mu\sigma\hat D^\mu\sigma+\hat R\sigma\}
\ee
After performing the integration over the ghost fields one is
led to
\be\label{1.6}
S[\phi;\hat g]=-\frac{\kappa^2}{32\pi}I[\hat g]+S_L[\phi;\hat g]\ee
with the Liouville action
\be\label{1.7}
S_L[\phi;\hat g]=\frac{\kappa^2}{8\pi}\int d^2x\sqrt{\hat g}\left\{
\hat D_\mu\phi\hat D^\mu\phi+\hat R\phi+\frac{m^2}
{2}e^{2\phi}\right\}\ee
and
\be\label{1.8}
3\kappa^2\equiv 26-c\ee
Here $\hat D_\mu$ and $\hat R$ are constructed from $\hat g_{\mu\nu}$.
The action $S[\phi;\hat g]$ coincides with $\Gamma_{\rm ind}[e^{2\phi}
\hat g]$ for $c$ replaced  by $c-26$. This substitution takes care
of the Faddeev-Popov determinant related to the conformal gauge
fixing. The parameter $m$ has dimension of mass,
$m^2\equiv 16\pi\lambda/\kappa^2$, and we assume $\kappa^2>0$
and $m^2>0$. What remains to be done is a functional integration over
$\phi$.\footnote{The subsequent integration over the moduli
implicit in $\hat g$ and the summation over the genus of the
Riemann surface is outside the scope of this investigation.}

In this paper we consider (\ref{1.7}) as the classical
action of a field  $\phi$ in a fixed background geometry $\hat g
_{\mu\nu}$, and we try to quantize $\phi$ using the exact
evolution equation. Because the decomposition (\ref{1.4}) is invariant
under the Weyl transformation
\be\label{1.9}
\phi'(x)=\phi(x)-\sigma(x),\quad \hat g_{\mu\nu}'(x)=e^{2\sigma(x)}
\hat g_{\mu\nu}(x)\ee
the quantization should respect this ``background split symmetry''.
This means that the functional integral over $\phi$ has to be performed
with the Weyl invariant measure \cite{dk},\cite{mm},\cite{mes},
\cite{dorew}, which derives from the following distance function in
the space fields:
\be\label{1.10}
ds^2_{\rm Weyl}=\int d^2x\sqrt{\hat g}\ e^{2\phi(x)}\ \delta\phi(x)^2\ee
This measure is different from the translation-invariant
measure which is used for usual scalar field
theories. The latter is based upon the
$\phi$-independent line element
\be\label{1.11}
ds^2_{\rm trans}=\int d^2x\sqrt{\hat g}\ \delta\phi(x)^2\ee
As discussed above, this means that the initial value
$\Gamma_\Lambda$ should obey the modified  Ward identity
related to Weyl invariance.

Our paper is organized as follows. In section 2 we show in
detail how the difference between the Weyl-invariant and the
translation-invariant quantization manifests itself at the
level of the exact renormalization group equations. We
derive the Ward identities resulting from Weyl invariance
and we use them to show that for $k\to0$ the renormalization
group flow converges towards a conformal field theory of
central charge  $26-c$. In section 3 we truncate the space
of actions and use the Ward identities in order to determine
the initial point of the evolution, $\Gamma_\Lambda$.
It is one of the typical features of Liouville theory that
in order to end up in the right universality class,
$\Gamma_\Lambda$ cannot be chosen equal to the classical
action $S_L$ (\ref{1.7}). In section 4 we derive
the truncated  evolution equations. Their solution
for $c\leq1$ is
discussed in section 5.
In order to estimate
the degree of precision which can be achieved with the
truncation used, we compare the solution of the evolution
equation to the constraints imposed by the Ward identities.
Section 6 contains a brief comparison between Liouville theory
and a free Feigin-Fuks theory. Section 7 is devoted to a
general study of renormalization group fixed points. We show
that the Liouville potential is singled out by the property that
it is an eigenvector of the renormalization group flow
linearized about the Gaussian fixed point. Our conclusions
are presented in section 8. The overall picture emerging from
this discussion is the conjecture that for low momenta
the effective action
$\Gamma_{k\to0}$ of quantum Liouville theory has the same form
as the classical action or at least very closely resembles
the latter. Within certain approximations we find that
$\Gamma$ equals $S_L$ of (1.9) with  a renormalized
parameter $m^2$.
If exact, this would imply important ``non-renormalization
theorems'', which go beyond the well-known statement \cite{dj}
that the effective potential is essentially the same
as the classical one.

\section{Weyl Invariant Measure and Ward Identities}
\setcounter{equation}{0}

Let $\chi(x^\mu)$ be an arbitrary scalar field in a
background metric $g_{\mu\nu}.$\footnote{From now on we
write $g_{\mu\nu}$ for $\hat g_{\mu\nu}$ and omit the
hats from $D_\mu$ and $R$.}
The derivation of the exact evolution equation \cite{wet} starts
from the scale-dependent generating functional
\bear\label{2.1}
e^{W_k[J;g]}&=&\int {\cal D}_g\chi\exp[-S[\chi;g]+\int d^2x\sqrt g
J(x)\chi(x)\nonumber\\
&&-\frac{1}{2}\int d^2x\sqrt g\chi(x)R_k(-D^2)\chi(x)]\ear
Here ${\D}_g\chi$ stands for either the Weyl or the translation
invariant measure. The last term in the square bracket of
(\ref{2.1}) is a diffeomorphism-invariant infrared cutoff.
The function $R_k(p^2)$ vanishes if the eigenvalues $p^2$ of
$-D^2=-D_\mu D^\mu$ are much larger than the cutoff $k$, and
it becomes a constant proportional to $k^2$ for $p^2\ll k^2$. The
precise shape of the function $R_k(p^2)$ is not important,
except that it has to interpolate monotonically between
$R_k(\infty)=0$ and $R_k(0)=Z_kk^2$. (The meaning of the constant
$Z_k$ will be explained later.) Often we shall write
\be\label{2.1a}
R_k(-D^2)=Z_kk^2C(-D^2/k^2)\ee
where $C$ is a dimensionless function with $C(0)
=1$ and $C(\infty)=0$. (In some cases it is sufficient
to use a constant mass like cutoff with $C\equiv1$.) Expanding
$\chi(x)$ in terms of eigenfunctions of $D^2$, we see that
in $W_k$ (\ref{2.1}) the high-frequency
modes with covariant momenta
$p^2\gg k^2$ are integrated out without any suppression
whereas the low-frequency modes with $p^2\ll k^2$ are
suppressed by a smooth, mass-type cutoff term $\sim k^2\chi^2$.

The effective average action $\Gamma_k[\phi;g]$ is defined
as the Legendre transform of $W_k[J;g]$ with the infrared
cutoff subtracted \cite{wet}, \cite{gau}:
\be\label{2.2}
\Gamma_k[\phi;g]=\int d^2x\sqrt g\phi(x)J(x)-W_k[J;g]-
\frac{1}{2}\int d^2x\sqrt g\phi(x)R_k(-D^2)\phi(x)\ee
Here $J=J[\phi]$ has to be obtained by inverting the relation
\be\label{2.3}
\phi(x)\equiv<\chi(x)>=[g(x)]^{-1/2}
\frac{\delta W_k[J;g]}{\delta J(x)}\ee
Our claim is that $\Gamma_k$ is a coarse-grained free
energy or scale-dependent effective action which interpolates
between the microscopic action $\Gamma_\Lambda$
and the usual effective
action $\Gamma$ for $k=\Lambda$ and $k=0$, respectively. The
equality $\Gamma_{k=0}=\Gamma$ is easy to establish: For $k=0$
the function $R_k(p^2)$ vanishes by construction for all
$p^2$, and the cutoff term disappears from (\ref{2.1})
and (\ref{2.2}). On the other hand, the limit
$\lim_{k\to\infty}\Gamma_k[\phi;g]$ is more subtle. To investigate
it, let us first note that $\Gamma_k$ obeys the following
integral equation \cite{inteq} which follows by inserting (\ref{2.2})
into (\ref{2.1}):
\bear\label{2.4}
&&e^{-\Gamma_k[\phi;g]}=\int{\cal D}_g\chi\exp\left[-S[\chi;g]+\int
d^2x\Bigl(\chi(x)-\phi(x)\Bigr)\frac{\delta\Gamma_k
[\phi;g]}{\delta\phi(x)}\right]\cdot\nonumber\\
&&\exp\left[
-\frac{1}{2}\int d^2x\sqrt g\Bigl(\chi(x)-\phi(x)\Bigr)R_k(-D^2)
\Bigl(\chi(x)-\phi(x)\Bigr)\right]\ear
If $R_k$ is consistent with all symmetries, there is a heuristic
argument that $\Gamma_{k\to\infty}$ equals the classical action:
For $k\to\infty$ the function $R_k(p^2)$ approaches $Z_kk^2$
for any value of $p$. In this limit the second exponential in
(\ref{2.4}),   $$\exp[-\frac{1}{2}Z_kk^2\int d^2x
\sqrt g(\chi-\phi)^2],$$       is proportional to a delta-functional
$\delta(\chi-\phi)$.  If the constant of proportionality is
field-independent, we get indeed $\Gamma_{k\to\infty}=S$. Whether
this is actually the case can be investigated by a saddle point
approximation of the path-integral in (\ref{2.4}). It becomes
exact in the limit $k\to\infty$. One finds that $\Gamma_{k\to
\infty}$ and $S$ differ by a fluctuation determinant in the heavy
mass limit:
\be\label{2.4a}
\Gamma_{k\to\infty}=S+\frac{1}{2}\lim_{k\to\infty}\ln \det
[S^{(2)}+Z_kk^2]\ee
($S^{(2)}$ denotes the Hessian of $S$.) In this limit, the
fluctuation determinant gives rise to at most a couple of
relevant and marginal operators such as $\phi\partial^2
\phi,\phi^2$ and $\phi^4$ in $d=4$, say. These are precisely
the terms which, in a renormalizable theory, one would have
included in $S$ anyhow. Thus we can say that $\Gamma_{k\to\infty}
=S$ holds true modulo a change of the bare parameters contained in
$S$. For finite $\Lambda$ typical corrections beyond a shift
in the bare parameters are suppressed by inverse powers of $\Lambda$.
If this is the case one
can ignore the difference between $\Gamma_\Lambda$
and $S$ for most practical purposes.

In Liouville theory the situation
turns out to be more involved. In order
to arrive at a quantum theory which is in the right universality
class, namely that of conformal field theories with central
charge $26-c$, some parameters in $\Gamma_{k\to\infty}$ have to
be fine-tuned to a particular value. The situation is similar
to the quantization of gauge theories. In order to obtain
massless gauge fields for $k\to0$, their mass at the UV cutoff
must assume a precisely defined value. In the case of a Weyl
invariant measure this will force us to use a $\Gamma_\Lambda$
which differs from $S$ in a well-defined way.

It is important to realize that in the renormalization
group framework the starting point of the
evolution, $\Gamma_\Lambda$, is the only place
where the difference between the Weyl and the translation invariant
measure enters. The evolution equation as such is the same
in both cases. It is derived by taking a $k$-derivative
of eq. (\ref{2.1}) and reexpressing the result in terms
of $\Gamma_k$ (\ref{2.2}). Without making any assumption
about ${\cal D}_g\chi$ one finds
\be\label{2.5}
\frac{\partial}{\partial t}\Gamma_k[\phi;g]=\frac{1}{2}
{\rm Tr}\left[\left(\Gamma_k^{(2)}[\phi;g]+R_k(-D^2)\right)^{-1}
\frac{\partial}{\partial t}R_k(-D^2)\right]\ee
Here $t\equiv \ln k$ is the renormalization group
``time'' and $\Gamma_k^{(2)}$ and $R_k$ are
differential operators in
the variable $x$, related to the Hessian
\be\label{2.6}
\Gamma_k^{(2)}(x,y)=[g(x)g(y)]^{-1/2}
\frac{\delta^2\Gamma_k}{\delta\phi(x)\delta\phi(y)}\ee
by
\be\label{2.7}
\Gamma_k^{(2)}(x,y)=\Gamma_k^{(2)}   \ \delta(x-y)g(y)^{-1/2}
\ee
As mentioned in the introduction, the quantization program
consists of the following two steps:
(i) Find the initial value $\Gamma_\Lambda$ for large
$\Lambda$ for the classical  theory under consideration.
(ii) Solve the evolution equation (\ref{2.5}) for this
initial condition and let $k\to 0$ in the
solution $\Gamma_k$.

Our tool to determine $\Gamma_\Lambda$ are the Ward
identities related to the Weyl transformation (\ref{1.9})
which we are going to derive now. We use the Liouville action
$S=S_L$ of (\ref{1.7}) in (\ref{2.1}), and we subject
both sides of eq. (\ref{2.1}) to the transformation
\be\label{2.8}
J'=e^{-2\sigma}J,\quad g'_{\mu\nu}=e^{+2\sigma}
g_{\mu\nu}\ee
The integration variable on the r.h.s. of (\ref{2.1}) is
denoted $\chi'$, and together with (\ref{2.8}) we transform
it according to
\be\label{2.9}
\chi'=\chi-\sigma\ee
Under the combined transformation the response of the
Liouville action is
\be\label{2.10}
S_L[\chi';g']=S_L[\chi;g]-\frac{\kappa^2}{8\pi}
\int d^2x\sqrt g\{D_\mu\sigma D^\mu\sigma+R\sigma\}
\ee
which follows from $(\sqrt gR)'=\sqrt g(R-2D^2\sigma)$ and
similar identities. By its very definition the Weyl invariant measure is
invariant under the simultaneous transformation of $\chi$ and
$g_{\mu\nu}$, but the translation invariant measure changes
by the well-known Jacobian \cite{polya}, \cite{dorew}
expressing the conformal anomaly:
\be\label{2.11}
{\cal D}_{g'}\chi'={\cal D}_g\chi\exp\left[\frac{\tau}
{24\pi}\int d^2x\sqrt g\{D_\mu\sigma D^\mu\sigma
+R\sigma\}\right]\ee
Here $\tau=0\ (\tau=1)$ for the Weyl (translation) invariant measure.
Though we are mainly interested in the Weyl invariant
quantization, we treat both cases in parallel. Thus we are
led to
\bear\label{2.12}
W_k[e^{-2\sigma}J,\ e^{2\sigma}g]&=&\frac{3\kappa^2+\tau}
{24\pi}\int d^2x\sqrt g\{D_\mu\sigma D^\mu\sigma+R\sigma\}
\nonumber\\
{}\nonumber\\
-\int d^2x\sqrt g J\sigma&+&\ln\int{\cal D}_g\chi\exp\Bigl\{-S_L[\chi;
g]+\int d^2x\sqrt gJ\chi\nonumber\\
{}\nonumber\\
&-&\frac{1}{2}\int d^2x\sqrt ge^{2\sigma}(\chi-\sigma)R_k(-
D^2_{g'})(\chi-\sigma)\Bigr\}\ear
Expanding to first order in $\sigma$ one arrives at
\bear\label{2.13}
&&2\frac{g_{\mu\nu}}{\sqrt g}\frac{\delta W_k[J;g]}
{\delta g_{\mu\nu}(y)}-2\frac{J(y)}{\sqrt g}
\frac{\delta W_k[J;g]}{\delta J(y)}=\frac{3\kappa^2+\tau}{24\pi}
R(y)-J(y)\nonumber\\
{}\nonumber\\
&&\qquad\qquad+<R_k(-D^2)\chi(y)>-<\chi(y)R_k(-D^2)\chi(y)>\nonumber\\
{}\nonumber\\
&&\qquad\qquad-{\rm Tr}[\hat R(y)<\chi\otimes\chi>]\ear
with the $y$-dependent operator (acting on $x$)
\be\label{2.14}
\hat R_k(y)=\frac{g_{\mu\nu}(y)}{\sqrt{g(y)}}
\frac{\delta}{\delta g_{\mu\nu}(y)}R_k(-D^2[g_{\mu\nu}(x)])\ee
Eq. (\ref{2.13}) can be reformulated as a statement about
$\Gamma_k$ by using its definition (\ref{2.2}) and exploiting the
familiar property of the Legendre transform that the connected
2-point function
\be\label{2.16a}
G_k(x,y)=<\chi(x)\chi(y)>-\phi(x)\phi(y)\ee
is the matrix inverse of $(\Gamma_k^{(2)}+R_k)
(x,y)$. The result is the following Weyl-Ward identity for
the effective average action:
\bear\label{2.15}
&&2\frac{g_{\mu\nu}}{\sqrt g}\frac{\delta\Gamma_k[\phi;g]}{\delta
g_{\mu\nu}(x)}-\frac{1}{\sqrt g}\frac{\delta\Gamma_k[\phi;g]}
{\delta\phi(x)}=-\frac{3\kappa^2+\tau}{24\pi}R(x)\nonumber\\
{}\nonumber\\
&&\qquad+<x|R_k\left(\Gamma_k^{(2)}+R_k\right)^{-1}
|x>+{\rm Tr}\left[\hat R_k(x)\left(\Gamma_k^{(2)}
(x)+R_k\right)^{-1}\right]\ear
This is a formula for the trace of the energy momentum
tensor derived from $\Gamma_k$. For fields satisfying the
equation of motion $\delta\Gamma_k/\delta\phi=0$ it
consists of an anomaly term proportional to the
curvature scalar, plus two additional pieces which are due
to the explicit symmetry breaking by the cutoff. The Ward
identity is consistent with the evolution equation in the sense
that if the solution $\Gamma_k$ of (\ref{2.5}) satisfies the Ward
identity for one value of $k$, it automatically does so also for
any other value. More precisely, this is true for exact
solutions of the evolution equations. Approximate solutions,
obtained for instance by truncating the space of all action
functionals, are not necessarily consistent with the Ward identity.
In this case the Ward identity is a powerful check for the
quality of the truncation: a necessary condition for an
approximation to be reliable is that (\ref{2.15}) holds at
the desired level of accuracy. \footnote{In the same spirit the
modified Slavnov-Taylor identities where used to control
truncations in Yang-Mills theory \cite{ehw}.}

When an arbitrary (classical or quantum) conformal field
theory with action $S$ and central charge $c[S]$ is
coupled to gravity the trace of its energy momentum
tensor
\be\label{2.16}
T^{\mu\nu}[S]\equiv\frac{2}{\sqrt g}
\frac{\delta S}{\delta g_{\mu\nu}}\ee
is (at least on shell) given by
\be\label{2.17}
T^\mu_\mu[S]=-\frac{c[S]}{24\pi}R\ +\ const\ee
The classical Liouville action (\ref{1.7}) satisfies this
condition with the central charge
\be\label{2.18}
c[S_L]=3\kappa^2=26-c\ee
Furthermore, by construction $R_k$ vanishes for $k\to0$.
Therefore the explicit symmetry-breaking terms in the
Ward identity are absent in this limit, and (\ref{2.15})
tells us that the evolution ends at a theory which
is conformally invariant and has central charge
\be\label{2.19}
c[\Gamma_{k\to0}]=26-c+\tau\ee
For the Weyl-invariant measure $(\tau=0)$ this is the
correct value. In order to investigate the implications of the
Ward identity for $\Gamma_{k\to\infty}$ we have to make
an ansatz for $\Gamma_k$. We shall discuss this in detail
in section 3.

Before closing this section we mention a slight generalization
of the Ward identity. Let us couple the Liouville field in
a Weyl-invariant way to a set of external fields $\psi_i(x)$
which are not to be quantized:
\be\label{2.20}
S_\psi[\chi;g,\psi_i]=\sum_i\int d^2x\sqrt g\psi_i(x)e^{2(1-
\Delta^0_i)\chi}\ee
The action $S_\psi$ is invariant under
\be\label{2.21}
\psi_i'=e^{-2\Delta^0_i\sigma}\psi_i\ee
together with (\ref{2.8}) and (\ref{2.9}). The
$\psi_i$'s could be some spin-0 primary fields of the underlying
conformal theory of matter and ghost fields. Their bare
dimensions are $(\Delta_i^0,\Delta^0_i)$. We shall investigate
how these dimensions change when the system is coupled to
quantized gravity (``gravitational dressing''). Replacing
$S\equiv S_L$ by $S_L+S_\psi$ in the functional integral
(\ref{2.1}), the r.h.s. of the Ward identity remains unchanged,
but on the l.h.s. we get an additional piece:
\be\label{2.22}
{\cal L}\Gamma_k[\phi;g,\psi_i]=-\frac{26-c+\tau}{24\pi}R(x)+
{\cal A}_k(x)\ee
In (\ref{2.22}) we defined
\be\label{2.23}
{\cal L}\ \equiv \ 2\frac{g_{\mu\nu}(x)}{\sqrt g}\frac{\delta}
{\delta g_{\mu\nu}(x)}-\frac{1}{\sqrt g}\frac{\delta}
{\delta\phi(x)}-2\sum_i\Delta^0_i\ \frac{\psi_i(x)}{\sqrt g}
\frac{\delta}{\delta\psi_i(x)}\ee
and
\be\label{2.24}
{\cal A}_k(x)\ \equiv\  <x|R_k(\Gamma_k^{(2)}+R_k)^{-1}|x>+{\rm Tr}
\left[\hat R_k(x)(\Gamma_k^{(2)}+R_k)^{-1}\right]\ee
If ${\cal A}_k(x)$ contains $\phi$-dependent terms,
the equation (\ref{2.22}),
even with $\phi$ and $\psi_i$ on shell, is not of the form (\ref{2.17})
and $\Gamma_k$ is not a conformal theory, of course. We shall
see that even for $k\to\infty$ such terms are actually present and
prevent us from identifying $\Gamma_{k\to\infty}$ with the
classical action.

The spacetime-integrated version of the Ward identity can be
cast in a very suggestive form. It is not difficult to
show that
\be\label{2.25}
2R_k+2\int d^2y\sqrt{g(y)}\hat R_k(y)=
\partial_tR_k+\eta_kR_k\ee
which expresses the fact that a Weyl transformation of
$R_k$ with a constant $\sigma$ can be partially compensated
for by a corresponding change of $k$. In (\ref{2.25}) we
introduced the anomalous dimension
\be\label{2.26}
\eta_{k}\equiv-\partial_t\ln Z_k\ee
By using (\ref{2.25}) in the integrated Ward identity
and by eliminating the $\partial_t R_k$ term via the evolution
equation (\ref{2.5}) one arrives at
\bear\label{2.27}
\partial_t \Gamma_k&=&\int d^2x\sqrt g{\cal
L}\Gamma_k+\frac{26-c+\tau}{24\pi}\int d^2x\sqrt gR\nonumber\\
{}\nonumber\\
&&-\frac{1}{2}\eta_{k}{\rm Tr}\left[R_k(\Gamma_k^{(2)}+
R_k)^{-1}\right]\ear
We shall see later on that the $\eta_{k}$ term is negligible
actually. Thus (\ref{2.27}) shows that (for $\phi$ on shell,
$\delta\Gamma_k/\delta\phi=0$) the driving force for the
evolution is the difference between $T_\mu^\mu[\Gamma_k]$ and
the anomaly term for the central charge $26-c$ (for $\tau=0$). This
makes it explicit that the evolution of $\Gamma_k$ will stop
at a conformal theory with the right central charge.

\section{Truncation and Initial Value}
\setcounter{equation}{0}

As it seems impossible to find exact solutions of the
renormalization group equation, we try to construct an approximate
solution by truncating the space of actions. The idea is to
project the renormalization group flow on a subspace of the space
of all action functionals. In our case this subspace has finite
dimension and is coordinatized by a finite number of generalized
couplings. Therefore the approximated trajectory
$\{\Gamma_k,k\in[0,\Lambda]\}$ is described
by several functions of $k$ which are the solution of a
coupled system of ordinary differential equations.
The difficult task is to guess (and to justify, in a second
step) a truncation which leads to an approximate trajectory
$\Gamma_k$ as close as possible to the true one. For scalar
theories in $d>2$ dimensions, usually dimensional arguments
can be used as a guide line.
As long as anomalous dimensions are small, the importance of field
monomials $\phi^2,\phi^4, \phi^6,...$, say, may be judged
on the basis of their canonical dimensions. In $d=2$ the
situation is more involved since $\phi$ is dimensionless
and at first sight all functions of $\phi$ seem equally
important. (Canonically they are all marginal.) In the
case at hand we shall make an ansatz for the truncated
$\Gamma_k$ which is inspired by the form of the classical
Liouville action. We solve the evolution equation
using the truncation  on the r.h.s.. We investigate
explicitly if the general form of our ansatz remains
approximately conserved by the flow. For this purpose
we consider on the l.h.s. an extended class of actions
with free functions instead of only a finite number of
couplings. As
a consistency check, we show that the resulting
trajectory satisfies the Weyl-Ward identities to a very
good approximation. More importantly, the Ward identity
determines the correct initial value $\Gamma_\Lambda$ at the
UV cutoff $\Lambda$.

We work within a general space of actions
consisting of three pieces,
\be\label{3.1}
\Gamma_k[\phi;g,\psi]=\Gamma_k^L[\phi;g]+
\Gamma_k^\psi[\phi;g,\psi]+
\Gamma_k^{\rm grav}[g],\ee
the most important one being
\be\label{3.2}
\Gamma_k^{\rm L}[\phi;g]=\frac{\kappa^2_k}
{8\pi}\int d^2x\sqrt g\Bigl\{\zeta_k(\phi)D_\mu
\phi D^\mu\phi+\omega_k(\phi)R+v_k(\phi)\Bigr\}\ee
Here the running parameter $\kappa_k$ is defined as the coefficient
of the $\phi R$-term
by the convention
\be\label{3.2a}
\frac{\partial \omega_k}{\partial\phi}(\phi=\phi_0)=1\ee
where the reference point
$\phi_0$ may be chosen conveniently, e.g. $\phi_0=0$.
The truncation which we shall use on the r.h.s. of
the flow equations and in the Ward identities
assumes an effective potential of the form
\be\label{3.2b}
v_k(\phi)=\frac{m_k^2}{2\alpha^2_k\kappa^2_k}e^{2\alpha_k\kappa_k
\phi}\ee
and
\be\label{3.2c}
\omega_k(\phi)=\phi,\quad \zeta_k(\phi)=\zeta_k\ee
In this case $\Gamma_k^L$ is parametrized by
four functions of $k$, namely $\kappa_k, \zeta_k$,
$m_k$ and $\alpha_k$, which have to be determined from
the evolution equation. The second piece,
\be\label{3.3}
\Gamma_k^\psi[\phi;g,\psi]=\frac{1}{16\pi}\sum_i
\left(\frac{m_{ik}}{\alpha_{ik}}\right)^2\int d^2x\sqrt g\
\psi_i\ \exp(2\alpha_{ik}\kappa_k\phi)\ee
is included in order to study the ``gravitational dressing''
of matter field operators $\psi_i(x)$.
It is a scale-dependent analogue
of $S_\psi$ in (\ref{2.20}). We shall determine the functions
$m_{ik}$ and $\alpha_{ik}$ from the evolution equation.
This will enable us to compute the ``dressed'' scaling
dimensions in presence of gravity \cite{kpz}, \cite{dav},
\cite{do}. The $\psi_i$'s are treated in the external field
approximation here. The third term in (\ref{3.1}) is the
$\phi$-independent pure
gravity action for which we concentrate on the piece
\be\label{3.4}
\Gamma_k^{\rm grav}[g]=-\frac{\tilde\kappa_k^2}
{32\pi}I[g]\ee
This is the $k$-dependent counterpart of the
term $-(\kappa^2/32\pi)I[\hat g]$ in the classical decomposition
(\ref{1.6}). Note that at the quantum level the functions
$\kappa_k$ and $\tilde\kappa_k$ are different in general.

For the evolution equation and the Ward identity we need
$\Gamma_k^{(2)}$, eq. (\ref{2.6}), which is essentially the
matrix of second functional derivatives with respect to
$\phi$. From (\ref{3.2}) with a $\phi$-independent coefficient
$\zeta_k$
we see that an eigenmode of $-D^2$ with
eigenvalue $p^2$ has the kinetic energy $(\zeta_k\kappa_k^2/4\pi)
p^2$. According
to the general rule \cite{wet}, \cite{gau} we identify
the constant $Z_k$ in the cutoff (\ref{2.1a}) with $\zeta_k
\kappa_k^2/4\pi$, i.e.
\be\label{3.8}
R_k(-D^2)=Z_kk^2C(-D^2/k^2)\ee
\be\label{3.8a}
Z_k=\frac{1}{4\pi}\zeta_k\kappa_k^2\ee
This makes sure that for the low momentum modes $(C
\approx 1)$ the kinetic term and the cutoff combine to
$Z_k(k^2+p^2)$, as it should be. In terms of $Z_k$ the
second functional derivative
reads
\be\label{3.5}
\Gamma_k^{(2)}=Z_k(-D^2+E_k)\ee
with
\be\label{3.6}
E_k=\tilde m_k^2\ \exp(2\alpha_k\kappa_k\phi)+\sum_i\tilde m
^2_{ik}\ \psi_i\ \exp(2\alpha_{ik}\kappa_k\phi)\ee
where
\be\label{3.7}
\tilde m_k^2=\frac{m_k^2\kappa_k^2}{4\pi Z_k}
=\frac{m^2_k}{\zeta_k}\quad,\quad
\tilde m_{ik}^2=\frac{m_{ik}^2\kappa_k^2}{4\pi Z_k}
=\frac{m_{ik}^2}{\zeta_k}.\ee

Next we derive the constraints which the Ward identities
impose on the functions $\kappa_k,\alpha_k,...$. This information
is needed to determine the correct initial value $\Gamma_\Lambda$.
In section 2 the pure gravity term was not included.
Hence $\Gamma_k$ of section 2 corresponds to $\Gamma_k^L+ \Gamma_k^\psi$.
Applying the differential operator
(\ref{2.23}) to the ansatz (3.2) with (3.4) and (3.5) yields
\bear\label{3.9}
{\cal L}(\Gamma_k^L+\Gamma_k^\psi)&=&\frac{\kappa_k^2}{4\pi}
(\zeta_k-1)D^2\phi-\frac{\kappa_k^2}{8\pi}R\nonumber\\
&&+\frac{1}{8\pi}\left(\frac{m_k}{\alpha_k}\right)^2
(1-\alpha_k\kappa_k)e^{2\alpha_k
\kappa_k\phi}\nonumber\\
&&+\frac{1}{8\pi}\sum_i\left(\frac{m_{ik}}{\alpha_{ik}}\right)^2
(1-\Delta^0_i-\alpha_{ik}\kappa_k)\psi_ie^{2\alpha_{ik}
\kappa_k\phi}\ear
Similarly ${\cal A}_k$ has to be projected on the
subspace of actions selected by the truncation. We make
an ansatz containing the same field monomials as (\ref{3.9})
and ignore all other invariants:
\bear\label{3.10}
{\cal A}_k&=&-\frac{\kappa_k^2}{4\pi}{\cal A}_k^{(1)}
\ D^2\phi+\frac{1}
{24\pi}{\cal A}_k^{(2)}\ R\nonumber\\
&&+\frac{1}{8\pi}m^2_k\ {\cal A}^{(3)}_k\ e^{2\alpha_k\kappa_k\phi}
\nonumber\\
&&+\frac{1}{8\pi}\sum_im^2_{ik}\ {\cal A}^{(4)}_k\ \psi_i\ e^{2\alpha
_{ik}\kappa_k\phi}+...\ear
In this manner the Ward identities boil down to the
following conditions which should hold for all values of
$k$:
\be\label{3.11}
\zeta_k=1-{\cal A}_k^{(1)}\ee
\be\label{3.12}
3\kappa_k^2=26-c+\tau-{\cal A}^{(2)}_k\ee
\be\label{3.13}
\alpha_k\kappa_k+\alpha_k^2{\cal A}_k^{(3)}-1=0\ee
\be\label{3.14}
\alpha_{ik}\kappa_k+\alpha_{ik}^2{\cal A}^{(4)}_k+\Delta_i^0-1=0\ee
In general the coefficients ${\cal A}_k^{(n)}$ depend on the choice
for the cutoff function $C$. In order to determine
the initial conditions for large $\Lambda$ we need their
values only for $k\to\infty$. In this limit all cutoffs
$C(-D^2/k^2)$ become equivalent to the
mass-like cutoff $C\equiv 1$. For later use we compute
the ${\cal A}_k^{(n)}$'s also for finite values of $k$, but for
simplicity we restrict ourselves to $C\equiv 1$.
Note that $\hat R_k\equiv 0$ in this case.
The terms proportional to the curvature scalar and to $D^2\phi$
are most easily extracted from ${\A}_k(x)$ by
writing
\be\label{3.18}
{\A}_k(x)=k^2\int^\infty_0dse^{-sk^2}<x|e^{s(D^2-E_k)}
|x>\ee
with $E_k(x)=\tilde m_k^2(1-2\alpha_k\kappa_k\phi+...)$ and
by using the standard heat kernel expansion
\be\label{3.19}
<x|e^{s(D^2+P(x))}|x>=\frac{1}{4\pi s}[1+\frac{1}{6}Rs+\frac{1}{6}
D^2Ps^2+...]\ee
Thus one finds
\be\label{3.15}
{\cal A}_k^{(1)}=\frac{\alpha_k}{3\kappa_k}\frac{k^2\tilde m^2_k}
{(\tilde m_k^2+k^2)^2}\ee
\be\label{3.16}
{\cal A}_k^{(2)}=\frac{k^2}
{\tilde m_k^2+k^2}\ee
In order to determine ${\A}_k^{(3)}$ and ${\A}^{(4)}_k$ we may set
$\phi,\psi_i=const,\  g_{\mu\nu}=\delta_{\mu\nu}$
such that
\be\label{3.20}
{\A}_k(x)=\int\frac{d^2p}{(2\pi)^2}\frac{k^2}{p^2+k^2+E_k}\ee
The result depends on the relative size of $k^2$ and $E_k$.
For $k^2\gg E_k$ we can expand in $E_k$ and find up to a constant
\be\label{3.24a}
{\cal A}_k=-E_k\int\frac{d^2p}{(2\pi)^2}\frac{k^2}{(p^2+k^2)^2}
=-\frac{m_k^2}{4\pi\zeta_k}\exp(2\alpha_k\kappa_k\phi)
-\sum_i\frac{m^2_{ik}}{4\pi\zeta_k}\psi_i\exp(2\alpha_{ik}\kappa_k
\phi)\ee
For this region, which is relevant for $k=\Lambda$, one can
extract
\be\label{3.24b}
{\cal A}_k^{(3)}=A^{(4)}_k=-2\zeta_k^{-1}\ee
On the other hand, for small $k$ we have to use the
general answer
\be\label{3.24c}
{\cal A}_k(\phi)-{\cal A}_k(\phi\to-\infty)=-\frac{k^2}{4\pi}
\ln\frac{k^2+E_k}{k^2}\ee
which vanishes for $k\to0$ as it should be. We infer ${\A}_{k\to0}
^{(3)}={\cal A}^{(4)}_{k\to0}=0$.

Let us now discuss what this implies for the initial point of
the evolution $\Gamma_\Lambda$. Letting $k\to\infty$
we obtain at the UV cutoff  $\Lambda$
\be\label{3.21}
{\A}_\Lambda^{(1)}=0,\quad {\A}_\Lambda^{(2)}=1,
\quad {\A}_\Lambda^{(3)}=
{\A}_\Lambda^{(4)}=-2\zeta_\Lambda^{-1}\ee
By (\ref{3.11})
with (\ref{3.21}) the initial wave function normalization constant
is fixed to be unity,
\be\label{3.21a}
\zeta_\Lambda=1\ee
such that ${\A}^{(3)}_\Lambda
={\A}_\Lambda^{(4)}=-2$. By (\ref{3.12}) the initial value for the
coefficient of the $R\phi$-term in the action
is
\be\label{3.22}
3\kappa_\Lambda^2=25-c+\tau\ee
In the classical action $S_L$ the corresponding coefficient
of $R\phi/24\pi$ is $c[S_L]=3\kappa^2=26-c$. For the translation
invariant quantization $(\tau=1)$ these values coincide, but
for the Weyl invariant case $(\tau=0)$ we have to start
from a different value, $3\kappa_\Lambda^2=25-c$, and
$\Gamma_\Lambda^L$ cannot coincide with $S_L$.
We emphasize that $3\kappa_\Lambda^2$ does not have
the interpretation of a central charge if $\Gamma_\Lambda$ does
not describe a conformal field theory. As we remarked at the
end of section 2, any $\phi$-dependence of ${\cal A}_\Lambda$ means
that the trace of the energy momentum tensor of $\Gamma_\Lambda$
does not have the form (\ref{2.17}). The nonzero values of
${\cal A}_\Lambda^{(3)}$ and ${\cal A}_\Lambda^{(4)}$
imply that for
$m_\Lambda,m_{i\Lambda}\not=0$ such terms are indeed present in
$T^\mu_\mu[\Gamma_\Lambda]$. In the exceptional case $m_\Lambda,
m_{i\Lambda}=0,\  \Gamma_\Lambda$ is a conformal free field theory
of the Feigin-Fuks type $\cite{ff}$ with central charge $25-c+\tau$.

From (\ref{3.13}) we also obtain a relation which determines the
initial value $\alpha_\Lambda$ in terms of $\kappa_\Lambda$,
\be\label{3.23}
\alpha_\Lambda\kappa_\Lambda=1+2\alpha^2_\Lambda\ee
With $\tau=0$ for the Weyl-invariant measure we have
the two solutions
\be\label{3.24}
\alpha_\Lambda=\frac{1}{4\sqrt3}[\sqrt{25-c}\pm\sqrt{1-c}]\ee
with the perturbative branch $(\alpha_\Lambda\kappa_\Lambda=
1+0(1/c)$ for $c\to-\infty)$, corresponding to the minus sign.
The starting value of the coefficient in the Liouville exponential
is fixed to be
\be\label{3.25}
\alpha_\Lambda\kappa_\Lambda=\frac{1}{12}[25-c\pm\sqrt{(c-25)
(c-1)}]\ee
It is real only for $c\leq 1$ and $c\geq25$. In the following
we restrict ourselves to $c\leq1$. Eq. (\ref{3.25}) is
familiar from other approaches to Liouville theory \cite{dorew}.
In the present context its interpretation is different, however.
Likewise eq. (\ref{3.14}) yields
\be\label{3.26}
2\alpha_{i\Lambda}^2-\alpha_{i\Lambda}\kappa_\Lambda-\Delta^0_i
+1=0\ee
which fixes $\alpha_{i\Lambda}$ in terms of $\Delta_i^0$ and $\kappa
_\Lambda$. Knowing $\alpha_{ik}$ and $\alpha_k$ at a certain
scale $k$ is equivalent to knowing the gravitationally dressed scaling
dimension $\Delta_i$ of $\psi_i$. By definition this dimension
describes the scaling of $\psi_i$ relative to the one of the
area operator which is governed by $\alpha_k$. One defines
\cite{dk}, \cite{dav} $\Delta_i(k)\equiv1-\alpha_{ik}/\alpha_k$,
and obtains from (\ref{3.26}) for $k=\Lambda$ and the
sign corresponding to the perturbative branch
\be\label{3.27}
\Delta_i(\Lambda)=\frac{\sqrt{\kappa^2_\Lambda-8+8\Delta^0_i}
-\sqrt{\kappa_\Lambda^2-8}}{\kappa_\Lambda-\sqrt{\kappa_\Lambda^2-8}}
\ee
Or, with $\tau=0$,
\be\label{3.28}
\Delta_i(\Lambda)=\frac{\sqrt{1-c+24\Delta^0_i}
-\sqrt{1-c}}{\sqrt{25-c}-\sqrt{1-c}}
\ee
The r.h.s. of this expression is precisely what appears in the famous
KPZ-formula \cite{kpz} for the scaling dimension of $\psi_i$ in presence
of quantized gravity. However, in the renormalization group
framework the properties of the quantum theory are obtained
for $k\to0$ rather than at $k=\Lambda$. Therefore (\ref{3.28})
is not the KPZ-formula yet. It obtains only provided we can
show that $\Delta_i(0)=\Delta_i(\Lambda)$.

Finally we have to fix the initial value $\tilde\kappa_\Lambda$
in the pure gravity piece (\ref{3.4}). This term was not
included when we derived the Ward identity in section 2
because it can be pulled in front of the path integral
(\ref{2.1}). ${\cal A}_{k\to\infty}$ produces no term proportional
to $I[g]$, so we may read off $\tilde\kappa_\Lambda$ from
the classical action (\ref{1.6}) with (\ref{1.8}):
\be\label{3.29}
3\tilde\kappa_\Lambda^2=26-c\ee
As $\kappa_\Lambda\not=\tilde\kappa_\Lambda$ for $\tau=0$, $\Gamma
_\Lambda^L$ and $\Gamma_\Lambda^{\rm grav}$ cannot be combined
into a functional of $g_{\mu\nu}e^{2\phi}$, and $\Gamma_\Lambda$
is clearly not Weyl-invariant. This lack of Weyl invariance
compensates for the one arising from the infrared cutoff term
$\Delta_\Lambda S$.
Weyl invariance is restored when all fluctuations
are integrated out at $k=0$.

Our method of fixing the initial condition for the
Weyl-invariant quantization has a certain similarity
with the approach of Distler and Kawai \cite{dk}
where an ansatz is made for the Jacobian relating the
Weyl- to the translation-invariant measure and where
Weyl-Ward identities (different from ours) are used to
fix the free parameters \cite{mes}. This similarity is
accidental, however, because we happend to use a truncation
for $\Gamma_k$ which is similar to the classical Liouville action.
Our method works for more general truncations in the same way.
The Ward identity picks the ``best'' initial point
from the subspace of truncated actions. It is not necessary
here to assume that the Jacobian has the classical Liouville form.

To summarize this section, the initial value $\Gamma_\Lambda$
is given by eq. (\ref{3.1}),
with $\zeta_\Lambda=1$, $3\kappa^2_\Lambda=25-c,\ 3\tilde
\kappa_\Lambda^2=26-c$ and $\alpha_\Lambda,\alpha
_{i\Lambda}$ specified by (\ref{3.25}) and (\ref{3.26}).
This holds for $\Lambda\to\infty$. Since $\Lambda$ has the dimension
of a mass, the precise meaning of the last statement involves
the ratio $\Lambda/m$. It will be specified in the next section.

\section{The Evolution Equations}
\setcounter{equation}{0}

In this section we derive the coupled system
of ordinary differential equations for the functions
$\alpha_k,m_k,...$ appearing in the truncation ansatz.
In order to investigate the universality properties of the
various $\beta$-functions, we employ an unspecified cutoff
function $C$. Upon inserting the truncation (\ref{3.1}),
its Hessian (\ref{3.5}) and the cutoff (\ref{3.8}) into
the renormalization group equation (\ref{2.5}) we have to
evaluate
\be\label{4.1}
\partial_t\Gamma_k=B_k-\frac{1}{2}\eta_kB_k^{(\eta)}\ee
with (a prime denotes the derivative with respect to the
argument):
\be\label{4.2}
B_k\equiv{\rm Tr}\biggl[\Bigl(k^2C(-D^2/k^2)+D^2C'(-D^2/k^2)\Bigr)
\ \Bigl(-D^2+k^2C(-D^2/k^2)+E_k\Bigr)^{-1}\biggr]\ee
\be\label{4.3}
B_k^{(\eta)}\equiv {\rm Tr}\biggl[k^2C(-D^2/k^2)\Bigl(-D^2+k^2
C(-D^2/k^2)+E_k\Bigr)^{-1}\biggr]\ee
Here
\be\label{4.4}
\eta_k\equiv-\partial_t\ln Z_k=-\partial_t\ln
(\kappa_k^2\zeta_k)\ee
is the anomalous dimension of the Liouville field.

The projected evolution equation (\ref{4.1}) is to be understood
as follows. Its l.h.s. is given by the ansatz (\ref{3.1})
with $\kappa_k^2$ replaced by $\partial_t\kappa_k^2$, etc..
When the r.h.s. is expanded in a basis of invariants constructed
from $\phi,g_{\mu\nu}$ and $\psi_i$, only the invariants also
present on the l.h.s. should be retained. The projection on the
subspace means that we discard all other terms. The evolution
equations for $\kappa_k^2$,... obtain by comparing the coefficients
of the respective invariants on both sides of eq. (\ref{4.1}).
The coefficient of the $R\phi$-term determines the running
of $\kappa_k^2$. Once it is known, the prefactor of $(D_\mu\phi)^2$
gives us the evolution of $\zeta_k$. This method is meaningful
only if the invariants are linearly independent in an appropriate
sense. This is certainly the case for the ``kinetic'' terms $(D
_\mu\phi)^2$ and $R\phi$. They are the first terms in an expansion
in the number of derivatives and powers of the curvature tensor,
and they could be extended to a complete basis in the full space
by adding all terms of the type $(D^2\phi)^2,R\phi^2,(D^2\phi)R\phi,
\cdots$.
For the potential $v_k(\phi)$ the situation is slightly more
complicated. In the spirit of a Laplace or a Fourier transform
we use the exponentials $\exp(\gamma\phi)$ as basis vectors and
consider them independent for different values of $\gamma$. In
the truncation only a single basis vector is retained, the one with
$\gamma=2\alpha_k\kappa_k$. This basis vector is allowed
to change with $k$. Finally we assume that the set $\{\psi_i\}$ is
linearly independent in the above sense and does not contain the
identity operator. Hence $\exp(2\alpha_k\kappa_k\phi)$
and $\psi_i\exp(2\alpha_{ik}\kappa_k\phi)$ are also
independent.

For a more general investigation we will occasionally
consider on the l.h.s. an enlarged ansatz with arbitrary functions
$v_k(\phi),\ \zeta_k(\phi)$ and $\omega_k(\phi)$ in eq. (\ref{3.2}). This
allows us to study to what extent invariants beyond our
truncation are generated by the evolution. It is in this
generalized context that we will see that
the exponential form of the potential does not change.

As we expect $\eta_k$ to be small, we ignore the term in (\ref{4.1})
proportional to $\eta_k$ in a first approximation and solve only
$\partial_t\Gamma_k=B_k$. This procedure will prove consistent,
because for the resulting solution $\eta_k$ is indeed
negligible. More precisely, we shall see that $Z_k$ varies
between $k=\Lambda$ and $k=0$ only by a finite amount even in
the limit when $\Lambda$ is sent to infinity. We will see below that
this change occurs for $k^2\approx \tilde m^2$ which shows that
$\eta_k$ vanishes both for $k^2\gg \tilde m^2$ and $k^2\ll
\tilde m^2$.

We start with the term $\sim \int d^2x\sqrt gR\omega
_k(\phi)$ generated
by $B_k$. Because it contains no $\psi_i$ and no derivatives
of $\phi$, we may set $\psi_i(x)\equiv 0$ and $\phi(x)=\phi=const$
for its evaluation. Then $E_k(x)$, (\ref{3.6}), becomes independent
of $x$. If we take advantage of the heat-kernel
expansion (\ref{3.19}),
we see that
\be\label{4.5}
{\rm Tr} [f(-D^2)]=\frac{1}{24\pi}\int d^2x\sqrt g Rf(0)+...\ee
for every function of the scalar Laplacian which can be written
as a Laplace or Fourier
transform:
$f(-D^2)=\int^\infty_0ds\tilde f(s)\exp(sD^2)$. As a result,
\be\label{4.6}
B_k=\frac{1}{24\pi}\int d^2x\sqrt g R\frac{k^2C(0)}{k^2C(0)+E_k}
\ + \ {\rm other \ invariants}\ee
Recall that by definition $C(0)$ equals unity for {\it any}
choice of the cutoff function. We see that the
coefficient of the $\sqrt gR$ term, and hence the $\beta$-function
for $\kappa^2_k$, is universal in the sense that it does not
depend on the form of the cutoff. More explicitly, we find
the flow equation (omitting occasionally the subscripts $k$ for the
couplings)
\be\label{4.6a}
\partial_t(\kappa^2\omega(\phi))=\frac{1}{3}\frac{k^2}{k^2+\tilde
m^2\exp(2\alpha\kappa\phi)}\ee
For example, an expansion around $\phi=0$ yields
\be\label{4.6b}
\partial_t(\kappa^2\omega(0))=\frac{1}{3}\frac{k^2}{k^2+
\tilde m^2}\ee
and,  with the reference point $\phi_0=0$ in (3.3),
\be\label{4.7}
\partial_t\kappa^2=\partial_t(\kappa^2\omega'(0))=-\frac{2}{3}
\alpha\kappa\frac{k^2\tilde m^2}{(k^2+\tilde m^2)^2}\ee
On the other hand, for $\phi$ much smaller than a critical
value $\phi_c$
\be\label{4.7a}
\phi_c(k)\equiv\frac{1}{2\alpha_k\kappa_k}\ln\frac{k^2}{\tilde
m^2_k}\ee
one has
\be\label{4.7b}
\partial_t\Bigl(\kappa^2\omega(\phi\ll\phi_c)\Bigr)=\frac{1}{3}
\Bigl(1-\frac{\tilde m^2}{k^2}\exp(2\alpha\kappa\phi)\Bigr)\ee
Qualitatively, one finds that $\kappa^2\omega(\phi)$
changes at a constant rate $\partial_t(\kappa^2\omega)
=\frac{1}{3}$ for $\phi\ll\phi_c$ whereas the evolution
stops due to suppression factors for $\phi\gg\phi_c$:
\be\label{4.7c}
\partial_t\left(\kappa^2\omega(\phi\gg\phi_c)\right)=\frac{1}{3}
\exp(-2\alpha\kappa(\phi-\phi_c))\ee

Next we consider the evolution of $\zeta_k(\phi)$ and
$v_k(\phi)$. These pieces can be extracted from the flow
of $\Gamma_k$ for $g_{\mu\nu}=\delta_{\mu\nu}$. This part of the
investigation is therefore completely parallel to a
two-dimensional scalar theory in flat space \cite{avact},
\cite{ng},\cite{mor} as given by
\be\label{4.7i}
\Gamma_k[\phi]=\int d^2x\left\{\frac{1}{2}Z_k(\phi)\partial_\mu\phi
\partial^\mu
\phi+U_k(\phi)\right\}\ee
with effective potential
\be\label{4.7j}
U_k(\phi)\equiv\frac{\kappa_k^2}{8\pi} v_k(\phi)\ee
and ``wave function renormalization''
\be\label{4.7k}
Z_k(\phi)\equiv\frac{\kappa_k^2}{4\pi}\zeta_k(\phi)\ee
We start with the evolution of the effective potential
$U_k$. In a first step we keep on the l.h.s. of the flow
equation an arbitrary function of $\phi$ whereas on the
r.h.s. the ansatz (\ref{3.2b})
\be\label{4.7l}
U_k(\phi)=\frac{\tilde m_k^2Z_k}{(2\alpha_k\kappa_k)^2}\exp
(2\alpha_k\kappa_k\phi)\ee
is inserted. This allows us to investigate if terms different
from (\ref{4.7l}) are generated
in the course of the evolution,
i.e., if the effective potential changes away from its form of a
simple exponential. For a computation of the flow $\partial_tU_k$
we set $\psi_i=0, \phi=const.$ Hence the trace over functions
of $D^2=\partial^2$ becomes a simple  momentum integral.
The evolution of $U_k$ is therefore
given by the standard flow equations for a two-dimensional scalar
theory \cite{avact}, \cite{ng}, \cite{mor}.
What is special, however, is our ansatz for
the potential on the r.h.s. for which one finds
\be\label{4.8}
\partial_tU(\phi)=\frac{k^2}{4\pi}\int^\infty_0dy\Bigl(C(y)-yC'(y)
\Bigr)
\Bigl[y+C(y)+\frac{\tilde m^2}{k^2}\exp(2\alpha\kappa\phi)
\Bigr]^{-1}\ee
The global behavior of $U$ can be understood by noting
that $C(y)-yC'(y)=\frac{\partial}{\partial k^2}
(k^2C(\frac{x}{k^2}))$ is positive for all $y=x/k^2$.
For a fixed value of $\phi$ therefore $U_k(\phi)$ always
decreases as $k\to0$. This decrease is faster for small $\phi$
than for large $\phi$ and the rate of decrease depends
monotonically on $\phi$. We see that $U_k(\phi)$ remains
monotonically increasing with $\phi$ in the whole range
of $\phi$ and can therefore not develop a minimum for
finite $\phi$! (The minimum remains at $\phi\to-\infty$).
Actually, this remark generalizes within the truncation
(\ref{3.2}) for arbitrary monotonically increasing potentials
with monotonically increasing second derivatives. Our finding
that quantum Liouville theory does not have a ground state
with finite constant $\phi$ seems therefore rather robust.
The more
quantitative behavior depends crucially on whether $\phi$
is smaller or larger than $\phi_c$: For $\phi\gg\phi_c$
the running is exponentially suppressed
\be\label{4.8a}
\partial_tU(\phi\gg\phi_c)=\frac{k^2}{4\pi}\exp
[-2\alpha\kappa(\phi-\phi_c)]\int^\infty_0
dy\Bigl(C(y)-yC'(y)\Bigr)\ee
In the opposite region $\phi\ll\phi_c$ we can
expand in the mass-like term
\be\label{4.8b}
\partial_tU(\phi\ll\phi_c)=
\frac{l_0}{4\pi}k^2-\frac{l_1}{4\pi}E_k=\frac{l_0}{4\pi}k^2
-\frac{\tilde m^2}{4\pi}\exp(2\alpha\kappa\phi)\ee
with
\be\label{4.8c}
l_0\equiv\int^\infty_0dy\frac{C(y)-yC'(y)}{y+C(y)}\ee
It is remarkable that the momentum integral\footnote{The integrals
$l_0, l_1$ belong to the standard integrals $l^d_n$
for $d=2$ appearing
in \cite{avact}, \cite{ng}.}
$(y\equiv p^2/k^2)$
\be\label{4.9}
l_1\equiv\int^\infty_0dy\frac{C(y)-yC'(y)}{(y+C(y))^2}=1\ee
is again independent of the detailed form of $C(y)$.
The integrand is a total derivative and it is sufficient to
know that $C(0)=1$ in order
to obtain the universal value $l_1=1$.
In consequence, for $\phi\ll\phi_c$ we find  that the evolution
generates in $U_k$ only a contribution proportional to
the coefficient multiplying $\exp(2\alpha\kappa\phi)$ plus
a constant term. Defining the constant
\be\label{4.10}
\lambda_L=U(\phi\to-\infty)\ee
one easily finds the solution
\be\label{4.10a}
\lambda_L(k)=\lambda_L(\Lambda)-\frac{l_0}{2\pi}(\Lambda^2-k^2)\ee
Subtracting $\lambda_L$ we see that for
$\phi\ll\phi_c$ the exponential shape
of $U$ does not change in the course of the evolution . This implies
$\partial_t(\alpha\kappa)
=0$, i.e., the product $\alpha_k\kappa_k$ is a
renormalization group invariant:
\be\label{4.11}
\alpha_k\kappa_k=const=\alpha_\Lambda\kappa_\Lambda\ee
As a result, (\ref{4.8b}) boils down to an evolution
equation for $\tilde m^2$
\be\label{4.12}
\partial_t\tilde m^2=\left[\eta-\frac{(\alpha\kappa)^2}{\pi}
\frac{1}{Z}\right]\tilde m^2=\left[\eta-\frac{4\alpha^2}
{\zeta}\right]
\tilde m^2\ee
We recall, however, that (\ref{4.11}) and (\ref{4.12})
become exact only for $\phi\ll\phi_c$. For $\phi\gg\phi_c$
the running of $U_k$ stops according to eq.
(\ref{4.8a}) and $\partial_t(Z
\tilde m^2)$ tends therefore to zero if the parametrization
(\ref{4.7l}) is used in this range of $\phi$.
This can be seen explicitly by applying a derivative
$\partial_t$ to the definition
\be\label{M1}
Z\tilde m^2=\frac{\kappa^2}{4\pi}m^2=\frac{\partial^2U}{\partial
\phi^2}(0)\ee
and by using (\ref{4.8}):
\bear\label{M2}
&&\partial_t(Z\tilde m^2)=-\frac{\alpha^2\kappa^2}{\pi}\tilde m^2
{\cal J}\left(\frac{\tilde m^2}{k^2}\right)\nonumber\\
&&{\cal J}\left(\frac{\tilde m^2}{k^2}
\right)\equiv\int^\infty_0
 dy(C-yC')\left(y+C-\frac{\tilde m^2}{k^2}\right)
\left(y+C+\frac{\tilde m^2}{k^2}\right)^{-3}\ear
For the special case $C=1$ this yields
\be\label{M3}
\partial_t(Z\tilde m^2)=-\frac{\alpha^2\kappa^2}{\pi}
\frac{k^4\tilde m^2}{(k^2+\tilde m^2)^2}\ee

If we treat the exponentials multiplied by $\psi_i$ in the
same manner we obtain in the region $\phi\ll\phi_c$
relations similar to (\ref{4.11})
and (\ref{4.12}) with $\alpha_k$ and $m_k$ replaced by
$\alpha_{ik}$ and $m_{ik}$, respectively. This means that
the gravitationally modified scaling dimension of $\psi_i$ does
not get renormalized in this region
\be\label{4.13}
\Delta_i(k)=1-\frac{\alpha_{ik}}{\alpha_k}=1-\frac{\alpha_{i\Lambda}}
{\alpha_\Lambda}=\Delta_i(\Lambda)\ee
Here the initial value $\Delta_i(\Lambda)$
is related to the classical
dimension $\Delta_i^0$ by eq. (\ref{3.28}), a consequence
of the Ward identity. Again, for $\phi\gg\phi_c$ the running
stops effectively.

In order to determine the flow of $Z_k$, we have to
extract the $(\partial_\mu\phi)^2$ piece from $B_k$, evaluated
for $g_{\mu\nu}=\delta_{\mu\nu},\ \psi_i=0$. Using
standard derivative expansion techniques,
 this can be done in complete analogy
to ref. \cite{avact}, \cite{ng} and one obtains\footnote{The
r.h.s. of eq. (\ref{A1}) corresponds to $-\tilde\eta Z_k$
in \cite{avact}, \cite{ng}.} $(x \equiv p^2)$
\be\label{A1}
\partial_t Z(\phi)=\frac{1}{8\pi}\left(\frac{\partial^3U}{\partial
\phi^3}\right)^2\int^\infty_0dx\ x\ \tilde\partial_t
\left\{\left(\frac{\partial P}{\partial x}\right)^2\left(
P+\frac{\partial^2U}{\partial\phi^2}\right)^{-4}\right\}\ee
where
\be\label{A2}
P(x) \equiv Z_kx+R_k(x)=Z_k\left[x+k^2C\left(\frac
{x}{k^2}\right)\right]\ee
and $\tilde\partial_t$ is a logarithmic $k$-derivative
acting only on $R_k$, i.e., $\tilde\partial_t=\partial_tR_k(\partial
/\partial R_k)$. For $\phi\ll \phi_c$ the running of
$Z(\phi)$ is suppressed by the exponential factor
$(\partial^3U/\partial\phi^3)=2\alpha\kappa\tilde m^2
Z\exp(2\alpha\kappa\phi)$ according to
\be\label{A3}
\partial_tZ(\phi\ll\phi_c)=-\frac{m_4}{\pi}
\alpha^2\kappa^2\frac{\tilde m^4}{k^4}\exp(4\alpha\kappa
\phi)\ee
where the constant $m_4$ is given by
\be\label{A4}
m_4=\int^\infty_0dy y^{-2}\frac{\partial}{\partial y}
\left\{(1+C'(y))^2y^4(y+C(y))^{-4}\right\}\ee
On the other hand, the running is also suppressed for $\phi
\gg\phi_c$ due to the large mass term $\partial^2U/
\partial\phi^2=\tilde m^2Z\exp(2\alpha\kappa\phi)$,
\be\label{A5}
\partial_tZ(\phi\gg\phi_c)=-\frac{2\alpha^2\kappa^2}{\pi}
\frac{k^4}{\tilde m^4}\exp(-4\alpha\kappa\phi)
\ \int^\infty_0dy\ y^2(1+C'(y))C''(y)\ee
For a fixed value of $\phi$ there is therefore
only a  small window in $k$, namely when $\phi_c(k)\approx
\phi$, for which $\partial_tZ(\phi)$ may differ
significantly from zero. This means that,
even for $\Lambda\to\infty$, the function $Z_k(\phi)$
changes between $k=\Lambda$ and $k=0$ by a finite
amount at most. As another
consequence, the anomalous dimension
$\eta$ turns out very small for all $k$ except possibly in the
window $k^2\approx \tilde m^2$.
In fact, from
\be\label{A6}
\eta=-\frac{1}{Z(0)}\partial_t Z(0)=-\frac{\alpha^2\kappa^2}
{2\pi}\tilde m^4 Z(0)\ \int^\infty_0 dx\ x\ \tilde \partial_t
\left\{\left(\frac{\partial P}{\partial x}\right)^2(P
+\tilde m^2Z)^{-4}\right\}\ee
one infers easily $\eta\sim\alpha^2\tilde m^4/k^4$ for $k^2
\gg\tilde m^2$ and $\eta\sim\alpha^2k^2/\tilde m^2$ for $k^2
\ll\tilde m^2$. For the special case $C(y)\equiv1$ one finds
\be\label{A7}
\partial_tZ=-\frac{\alpha^2\kappa^2}{3\pi}
\frac{\tilde m^4k^2}{(k^2+\tilde m^2)^3}\ee
where $Z\equiv Z_k\equiv Z_k(\phi=0)$.

\section{Flow of the Effective Liouville Action}
\setcounter{equation}{0}

Next we solve the evolution equations derived in
the previous sections.  First we consider values of $\phi$
around zero and the limit $\tilde m^2\ll k^2$. In this
limit the couplings $\kappa, \alpha\kappa, Z$
and $\zeta$ remain constant and we can use on the r.h.s. of
the flow equations
\be\label{5.1}
\kappa_k=\kappa_\Lambda,\ \alpha_k\kappa_k=\alpha_\Lambda\kappa
_\Lambda,\ Z_k=\frac{\kappa_\Lambda^2}{4\pi},\ \zeta_k=1\ee
The only nontrivial running concerns the ``mass parameter''
$\tilde m_k^2=m_k^2$ which is governed by an anomalous
mass dimension,
\be\label{5.2}
\partial_tm^2=-4\alpha_\Lambda^2m^2\ee
and similarly for $m_i^2$:
\be\label{5.3}
\partial_tm^2_i=-4\alpha_{i\Lambda}^2m_i^2\ee
The solution
\be\label{5.4}
m^2_k=\tilde m_k^2=\left(\frac{\Lambda}{k}\right)
^{4\alpha^2_\Lambda}m^2_\Lambda\quad,\quad
m_{ik}^2=\left(\frac{\Lambda}{k}\right)^{4\alpha^2_{i\Lambda}}
m^2_{i\Lambda}\ee
leads to an increase of $m^2_k$ and of the ratio $\tilde m_k^2/k^2$
as $k$ is lowered. There is therefore necessarily a value of $k$
for which the approximation $\tilde m^2/k^2\ll 1$ breaks down.
The ratio $\tilde m^2/k^2$ is the
only relevant parameter appearing in the various threshold
functions in the evolution equations. We conclude that for
values $k^2\approx \tilde m_k^2$ there is a qualitative change
in the behavior where the running of $\tilde m_k^2$ stops.

For a more detailed investigation of the ``threshold behavior''
at $\tilde m^2/k^2\approx1$ we start with the ``wave function
renormalization'' $Z_k$. Since the anomalous dimension $\eta$ is
effectively a function of $\tilde m^2/k^2$ which differs
from zero only for $\tilde m^2/k^2\approx1$ and vanishes both
for $\tilde m^2/k^2\to0$ and $\tilde m^2/k^2\to\infty$
there can be only a finite shift in $Z_k$ as $k$ crosses
the threshold. Also $\alpha_k\kappa_k$ is effectively
a function $\tilde m^2/k^2$,
\be\label{5.5}
\frac{\alpha_k\kappa_k}{\alpha_\Lambda\kappa_\Lambda}
={\cal G}\left(\frac{\tilde m^2}{k^2}\right),\
{\cal G}(0)=1\ee
and the  solution of (\ref{A7}) takes the general form
\bear\label{5.6}
&&Z_k=\frac{\kappa_\Lambda^2}{4\pi}\left[1+\delta_Z\left(\frac{
\tilde m^2}{k^2}\right)\right]\nonumber\\
&&\delta_Z(0)=0,\quad \delta_Z\left(\frac{\tilde m^2}{k^2}\to
\infty\right)=\bar\delta\ear
for some constant $\bar\delta$.
We can now insert (\ref{5.5}) and (\ref{5.6}) into the
flow equation (\ref{M2}) for $\tilde m^2$:
\be\label{5.7}
\partial_t\tilde m^2=-4\alpha_\Lambda^2{\cal K}
\left(\frac{\tilde m^2}
{k^2}\right)\tilde m^2\ee
The new threshold function ${\cal K}(\tilde m^2/k^2)$ obeys
\bear\label{5.8}
&&{\cal K}\left(\frac{\tilde m^2}{k^2}\right)=\frac{{\cal G}^2
(\tilde m^2
/k^2){\cal J}(\tilde m^2/k^2)}{1+\delta_Z(\tilde m^2/k^2)}
-\frac{\eta(\tilde m^2/k^2)}{4\alpha_\Lambda^2}\nonumber\\
&&{\cal K}(0)=1,\quad {\cal K}\left(\frac{\tilde m^2}{k^2}\to\infty\right)=0\ear
The precise threshold behavior depends on the choice of the
infrared cutoff function $C$ but the stopping of the flow
of $\tilde m^2$ for $k^2\ll\tilde m^2$ is general.

The solution of the flow equation (\ref{4.7})
for $\kappa$
can be written in the form
\be\label{5.9}
\kappa_k^2=\kappa_\Lambda^2+\frac{1}{3}\int^{\Lambda^2}_{k^2}
dk^2\frac{\alpha\kappa\tilde m^2}{(k^2+\tilde m^2)^2}
\ee
The integral on the r.h.s. is convergent for
$\Lambda\to\infty$ and leads to a finite shift
in $\kappa$ as $k$ crosses the threshold at $k^2\approx\tilde m^2$.
Setting $w=\tilde m^2/k^2$ and defining $h(w)$ by
\be\label{5.10}
\frac{dh}{dk^2}=\frac{1}{k^2}\frac{{\cal G}(w)w}{(1+w)^2}\ee
one has
\be\label{5.11}
\kappa_k^2=\kappa_\Lambda^2+\frac{\alpha_\Lambda\kappa_\Lambda}{3}
\left[h\left(\frac{\tilde m_\Lambda^2}{\Lambda^2}\right)-h
\left(\frac{\tilde m_k^2}{k^2}\right)\right]\ee
The l.h.s. of the
differential equation (\ref{5.10}) for $h$ involves the
threshold function ${\cal K}(w)$ of (\ref{5.8}) according to
\bear\label{5.12}
\frac{dh}{dk^2}&=&\frac{dh}{dw}\left(\frac{1}{k^2}
\frac{\partial\tilde m^2}{\partial k^2}-\frac{\tilde m^2}{k^4}\right)
\nonumber\\
&=&-\frac{dh}{dw}\left(2\alpha_\Lambda^2{\cal K}(w)w+w\right)\frac{1}{k^2}
\ear
Hence (5.10) is equivalent to
\be\label{5.13}
\frac{dh}{dw}=-\frac{1}{(1+2\alpha^2_\Lambda)(1+w)^2}{\cal H}
(w)\ee
Here we have collected all details of the threshold functions in
the factor
\be\label{5.14}
{\cal H}(w)\equiv\frac{(1+2\alpha_\Lambda^2){\cal G}(w)}
{1+2\alpha_\Lambda^2{\cal K}(w)}\ee
If we neglect these details and approximate ${\cal H}(w)=1$ the
integration of (\ref{5.13}) is easily done and leads to
\be\label{5.15}
\kappa_k^2=\kappa_\Lambda^2+\frac{1}{3}
\frac{\alpha_\Lambda\kappa_\Lambda}
{1+2\alpha^2_\Lambda}\left[\frac{\tilde m^2_k}{k^2+\tilde m^2_k}-
\frac{\tilde m^2_\Lambda}{\Lambda^2+\tilde m^2_\Lambda}\right]\ee
The second term in the bracket vanishes for $\Lambda\to\infty$. Also,
we may use $\alpha_\Lambda\kappa_\Lambda=1+2\alpha_\Lambda^2$
from the Ward identity (3.30).
We see that $\kappa_k^2$ increases by 1/3 as $k$ moves through
the threshold
\be\label{5.16}
\kappa_k^2=\kappa_\Lambda^2+\frac{1}{3}\frac{\tilde m_k^2}
{k^2+\tilde m^2_k}\ee
and,  for $k=0$, we end up at
\be\label{5.17}
\kappa_0^2=\kappa_\Lambda^2+\frac{1}{3}=\frac{26-c}
{3}\ee
This is exactly the result we had hoped to find
for a consistent truncation, namely that $3\kappa_k^2$
moves from $25-c$ for $k=\Lambda$ to $26-c$ for $k=0$.
A check if the details of the threshold functions do
not spoil this picture is interesting but nontrivial.
It requires a proof of the identity
\be\label{5.18}
\int^\infty_0dw\frac{{\cal H}(w)}{(1+w)^2}=1\ee

As for the wave function renormalization, the
Ward identities require that $\zeta_{k\to 0}=1$
which, according to eq. (5.6),
amounts to
\be\label{5.19}
\bar\delta=\frac{1}{25-c}=\frac{1}{3\kappa_\Lambda^2}\ee
To check whether this is the case
for our solution we combine eqs. (4.9) and (4.36) and find
\be\label{5.19a}
\frac{d\zeta}{dk^2}=\frac{\alpha^2}{3}
\left[ \frac{\zeta}{\alpha\kappa}\frac{\tilde m^2}
{(k^2+\tilde m^2)^2}-\frac{2\tilde m^4}{(k^2+\tilde m^2)^3}
\right] \ee
Approximating the running of $\tilde m_k^2$ by eq. (\ref{5.4}),
neglecting the evolution of $\alpha_k$ and using $\zeta_k/(\alpha
_k\kappa_k)\approx1$ (note $(\alpha\kappa)_{k\to0}=1$,
see below) one obtains for $\Lambda\to\infty$
\be\label{5.19c}
\zeta_k=\zeta_\Lambda-\frac{\alpha_\Lambda}{3\kappa_\Lambda}
\left[\frac{\tilde m^2_k}{k^2+\tilde m^2_k}-
\frac{\tilde m^4_k}{(k^2+\tilde m^2_k)^2}\right]\frac{\alpha_k^2}
{\alpha_\Lambda^2}\ee
For $k\to0$ the two terms in the bracket cancel and one indeed
obtains in this approximation
\be\label{5.19d}
\zeta_0=\zeta_\Lambda=1\ee
We remark that the approximations leading to both
${\cal H}(w)=1$ and $\zeta_0=1$ become exact in the leading
order of the semiclassical $(1/c)$ expansion.

In order to understand the shape of the potential
$U_k(\phi)$ for $k\to0$ we have to reconcile the running
of the prefactor of $\exp(2\alpha_\Lambda\kappa_\Lambda\phi)$
for $\phi<\phi_c$ with the approximate $k$-independence
of $U_k(\phi)$ for $\phi>\phi_c$. Expressed in different
terms, we have to incorporate the threshold function
in the running of the $\phi$-dependent mass term $\tilde m^2
\exp(2\alpha\kappa\phi)$ according to (\ref{M3}), (\ref{5.7}).
We will approximate
here the threshold effects by a sharp threshold and make the
approximation
\be\label{5.20}
U_k(\phi)=\frac{Z_\Lambda\tilde m_k^2(\phi)}
{4\alpha^2_\Lambda\kappa^2_\Lambda}\exp(2\alpha_\Lambda
\kappa_\Lambda\phi)\ee
with
\be\label{5.21}
\tilde m_k^2(\phi)=\left\{\begin{array}{ccc}
\mu_0k^{-4\alpha_\Lambda^2}& {\rm for}& \phi<\phi_c(k)\\
\mu_0k_c(\phi)^{-4\alpha_\Lambda^2}& {\rm for}& \phi>\phi_c(k)
\end{array}\right.\ee
and
\be\label{5.25}
\mu_0=\tilde m^2_\Lambda\Lambda^{4\alpha_\Lambda^2}=
m_\Lambda^2\Lambda^{4\alpha_\Lambda^2}\ee
Here $k_c(\phi)$ is the $\phi$-dependent scale where
the running stops as $\phi_c(k)$ becomes smaller than
$\phi$ for $k<k_c$.
It is given by the relation
\be\label{5.21a}
\phi=\frac{1}{2\alpha_\Lambda\kappa_\Lambda}\ln\frac{k_c^2}
{\tilde m_k^2(k_c)}\ee
with $\tilde m_k^2(k_c)=\mu_0k_c^{-4\alpha_\Lambda^2}$. This
yields
\be\label{5.22}
k_c(\phi)^{-4\alpha_\Lambda^2}=\Biggl[\mu_0\exp(2\alpha_\Lambda\kappa
_\Lambda\phi)\Biggr]^{-\frac{2\alpha^2_\Lambda}
{1+2\alpha_\Lambda^2}}\ee
In consequence, the potential for $\phi\gg\phi_c$ becomes
independent of $k$ and approaches the limit
\be\label{5.23}
U(\phi)=\frac{Z_\Lambda}{4\alpha_\Lambda^2\kappa^2_\Lambda}\
\mu_0^{\frac{1}{1+2\alpha_\Lambda^2}}\ \exp\left(
\frac{2\alpha_\Lambda\kappa_\Lambda}{1+2\alpha_\Lambda^2}
\phi\right)\ee
It is remarkable that the combination $\alpha_\Lambda\kappa
_\Lambda/(1+2\alpha_\Lambda^2)$ equals exactly unity by virtue of
the Ward identity at $k=\Lambda$. One concludes that the
effective exponent obeys
\be\label{5.24}
\alpha_k\kappa_k=1\quad {\rm for}\quad k\ll k_c\ee
where $k_c=\tilde m$ if
$\phi=0$.
The effective potential for the quantum Liouville
theory has the same exponential dependence as the classical
potential,
\be\label{5.26}
U(\phi)=\frac{\kappa_0^2\bar m^2}{16\pi}\exp(2\phi)\ee
but with a modified mass parameter
\be\label{5.27}
\bar m^2=\left(m^2_\Lambda\Lambda^{4\alpha_\Lambda^2}\right)^
{\frac{1}{1+2\alpha_\Lambda^2}}
\left(\frac{\alpha_0}{\alpha_\Lambda}\right)^2\ee

Similarly, the term in the effective Lagrangian which
is linear in the matter fields $\psi_i$ can be written in the form
$\sqrt g\sum_i H_i(\phi)\psi_i$, with
\be\label{5.28a}
H_i(\phi)=\frac{\mu_iZ_\Lambda}{4\alpha_{i\Lambda}^2\kappa
_\Lambda^2}\exp(2\alpha_{i\Lambda}\kappa_\Lambda\phi)
\left\{\begin{array}{lll}
k^{-4\alpha_{i\Lambda}^2}&{\rm for}& \phi<\phi_c(k)\\
k_c(\phi)^{-4\alpha_{i\Lambda}^2}&{\rm for}& \phi>\phi_c(k)\end
{array}\right.\ee
The reason is that if one expands (4.2) to first order in
$\psi_i$ one finds that the running of $\tilde m_{ik}$, too,
stops at $k\approx k_c$. (The threshold for $\tilde m_{ik}$
is determined by $\tilde m_k$, not by $\tilde m_{ik}$ itself.)
With (5.23) this yields for $\phi\gg\phi_c(k)$
\be\label{5.28b}
H_i(\phi)\sim\exp[(2\alpha_{i\Lambda}  \kappa_\Lambda-4
\alpha_{i\Lambda}^2  )\phi]\ee
or, with eq. (\ref{3.26}),
\be\label{5.28c}
\lim_{k\to0}\alpha_{ik}\kappa_k=\alpha_{i\Lambda}\kappa_\Lambda
-2\alpha_{i\Lambda}^2=1-\Delta_i^0\ee
In consequence, $\Delta_i(k)$ shifts away from $\Delta_i(\Lambda)$
as $k\to0$, where it approaches again the ``classical'' value
appearing in (2.23):
\be\label{5.28d}
\Delta_i(0)=1-\frac{\alpha_{i0}\kappa_0}{\alpha_0\kappa_0}
=\Delta_i^0\ee
In view of the KPZ relation this is certainly a surprising
result and we will comment on it in the conclusions.

The actions $\Gamma_k^L$ and $\Gamma_k^\psi$
are completely determined now. In order to judge the quality
of the truncation we can check how well
$\Gamma_k^L+\Gamma_k^\psi$
satisfies the Weyl-Ward identities of section 3. (Recall
that $\Gamma_k^{\rm grav}$ was not included there.)
As we mentioned already, the consistency of evolution equation
and Ward identity is automatic for the exact solution,
but it might be spoiled  if we use a ``bad'' truncation. Let
us insert the solutions for $m_k^2,...$ into (\ref{3.9})
and let us also use the conditions (\ref{3.23}) and (\ref{3.26})
which constrain the allowed initial conditions. We will
concentrate on the two regions $\phi\ll\phi_c$ and $\phi
\gg\phi_c$ and omit the more complicated threshold for
$\phi$ near $\phi_c$. (We have not solved the evolution
equation near the threshold with high precision either.)
Above the threshold, for $\phi\ll\phi_c$ or large $k$,
we find upon inserting our solution
\bear\label{5.28e}
&&\int d^2x\sqrt g{\cal L}(\Gamma_k^L+\Gamma_k^\psi)
+\frac{26-c+\tau}{24\pi}\int d^2x\sqrt gR\nonumber\\
&&=-\frac{1}{4\pi}m^2_k\left(\frac{\alpha_\Lambda}{\alpha_k}
\right)^2\int d^2x\sqrt ge^{2\alpha_\Lambda\kappa_\Lambda\phi}
\nonumber\\
&&\qquad-\frac{1}{4\pi}\sum_i m^2_{ik}
\left(\frac{\alpha_{i\Lambda}}{\alpha_{ik}}
\right)^2\int d^2x\sqrt g\ \psi_i\ e^{2\alpha_{i\Lambda}\kappa_\Lambda\phi}
\nonumber\\
&&=\partial_t(\Gamma_k^L+\Gamma^\psi_k)\ear
This is precisely the exact global (integrated)
Ward identity (\ref{2.27}) with $\eta_k=0$. This is a
very gratifying result which shows that our approximate
renormalization group trajectory should not be too far
away from the exact one in this region. The local Ward
identity contains more information than the global one.
For our truncation it is encoded in the equations (3.15)
-(3.18).
For $\tau=0$, eq. (3.16) with (3.22) yields
\be\label{R1}
3\kappa^2=26-c-\frac{k^2}{k^2+\tilde m^2}\ee
On the other hand, since for large $k$ the threshold function
${\cal H}(w)$ equals unity, the solution of the flow
equation is given by (\ref{5.16}):
\be\label{R2}
3\kappa^2=25-c+\frac{\tilde m^2}{k^2+\tilde m^2}\ee
This is exactly the same expression as (\ref{R1}).
Similarly, for $\alpha_k\to \alpha_\Lambda,\kappa_k\to\kappa_\Lambda$
the result of the flow equation for $\zeta_k$, eq. (\ref{5.19c}),
coincides with the Ward identity (3.15), (3.21).
We also note that for large $k$ the running of $\zeta, \kappa$
$\alpha\kappa$ and $\alpha_i\kappa$ can be neglected and the Ward
identities (\ref{3.13}), (3.18) are therefore obeyed.

Let us also turn to the Ward identities for $k=0$. Since
${\cal A}_k^{(n)}=0$, eqs. (3.15)-(3.18)
imply
\be\label{R3}
\zeta_0=1,\quad 3\kappa_0^2=26-c,\quad
\alpha_0\kappa_0=1,\quad \alpha_{i0}\kappa_0=1-\Delta_i^0\ee
These are exactly the values we have found from the solution
of the flow equation. In particular, we note that the Ward
identities (\ref{3.13}), (3.18) imply
$\Delta_i(0)=1-\alpha_{i0}\kappa_0/\alpha_0\kappa_0=
\Delta_i^{(0)}$, a result that we have already found from the
solution of the truncated evolution equation. We conclude that our
truncation seems to work very well, both for large and for small
$k$.

Before closing this section, we would like to comment on the use of
a running $Z_k$ in the definition of the IR-cutoff $R_k$ (2.2).
We have learned that $Z_k$ is only
shifted by a finite amount once $k$ drops below the threshold
value. In this particular case it can also be justified to
use a constant $Z_\Lambda$ instead of a running $Z_k$ in the
definition (2.2). In this version the last term $\sim\eta$
in the global Ward identity (2.30) is absent. Similarly, there
is no second term $\sim B_k^{(\eta)}$ in the evolution equation
(\ref{4.1}). The price to pay is a somewhat more complicated
momentum dependence of the propagator for $k$ below the threshold
which is due to the finite shift of the coefficient of the
$(D_\mu\phi)^2$ term. Since the running essentially stops in this
region, this actually is not a very severe disadvantage.

\section{Liouville versus Feigin-Fuks Theory}
\setcounter{equation}{0}

So far we have not yet determined the function
$\tilde\kappa_k$ which appears in the pure gravity action (\ref{3.4}).
For $C\equiv 1$, say, its $\beta$-function is obtained
by extracting the term proportional to the induced gravity
action $I[g]$ from the r.h.s. of
\be\label{6.1}
\partial_t\Gamma_k^{\rm grav}[g]={\Tr}[k^2(-D^2+k^2+E_k)^{-1}]\ee
We may set $\psi_i=0$, $\phi=0$ so that $E_k=\tilde m_k^2$.
The induced gravity action $I[g]$ arises by integrating
out massless fields. Massive fields would not give rise to
structures as nonlocal as $R\Delta_g^{-1}R$. Therefore the r.h.s.
of (\ref{6.1}) does not contain a piece proportional to $I[g]$ and
$\partial_t\tilde\kappa_k=0$. (This can be verified with the
techniques of ref. \cite{bv}.) Thus
\be\label{6.2}
3\tilde\kappa_k^2=3\tilde\kappa_\Lambda^2=26-c\ee
As a consequence, $\Gamma_k^{\rm grav}$ contributes a piece
${\cal L}\Gamma_k^{\rm grav}=(26-c)R/(24\pi)$ to the conformal anomaly,
and the global Ward identity for the complete action assumes the
form
\be\label{6.3}
\partial_t\Gamma_k-\int d^2x\sqrt g\ {\cal L}\Gamma_k=
\frac{\tau}{24\pi}\int
d^2x\sqrt gR\ee
For the Weyl-invariant measure this equation is equivalent to the
statement that $\Gamma_k$ is Weyl-invariant under the same
transformations as the classical action, together with
$k'=ke^{-\sigma}$, whereby only $x^\mu$-independent $\sigma$'s
are allowed:
\be\label{6.4}
\Gamma_{k'}[\phi';g_{\mu\nu}',\psi_i']=\Gamma_k[\phi;g_{\mu\nu},
\psi_i]\ee
In (\ref{6.4}) there is no anomaly term present.
This is as it should
be: When $k$ approaches zero, $\Gamma_k$ is heading for a conformal
field theory of the combined ghost + matter + Liouville system
whose total central charge is zero. Note also that $\tilde\kappa_k
=\kappa_k$ and $\zeta_k=1$ is precisely what is needed to combine
the $(D_\mu\phi)^2$ and $R\phi$-terms with the pure gravity
term into a manifestly Weyl-invariant action $\sim I[g_{\mu\nu}
e^{2\phi}]$.

In the above discussion we made the tacit assumption that
$m_\Lambda\not=0$. At this point a brief comparison of
Liouville theory with $m_\Lambda\not=0$ and the free
field case $m_\Lambda=0$ is in order. If we start the
evolution at a theory of the form
\be\label{6.5}
\Gamma_\Lambda^L(m=0)=\frac{\kappa_\Lambda^2}{8\pi}\int d^2
x\sqrt g\{D_\mu\phi D^\mu\phi+R\phi\}\ee
then the Ward identities still tell us that
$3\kappa_\Lambda^2=25-c$ and $3\tilde\kappa_\Lambda^2=26-c$.
(We set $\tau=0$ and $\psi\equiv0$ here.) The action (\ref{6.5})
is a Feigin-Fuks \cite{ff} conformal field theory of a free
field with a background charge whose nonminimal gravitational
interaction shifts the central charge away from unity.
Obviously (\ref{6.5}) is invariant under the renormalization group
evolution so that this time
\be\label{6.6}
3\kappa_{k=0}^2=3\kappa_\Lambda^2=25-c\quad(m_\Lambda=0)\ee
For $m_\Lambda=0$ it is not true any more that the
piece $\sim I[g]$ in
$\Gamma_k^{\rm grav}$ is a constant. The r.h.s. of (\ref{6.1}) can
be replaced by $\frac{1}{2}\partial_t{\Tr}\ln[-D^2+k^2]$
now, and this integrates to
\be\label{6.7}
\Gamma_{k=0}^{\rm grav}[g]=\Gamma_\Lambda^{\rm grav}
[g]+\frac{1}{2}{\Tr}
\ln[-D^2]=-\frac{3\tilde\kappa^2_\Lambda-1}{96\pi}I[g]\ee
because ${\Tr}\ln[-D^2]=I[g]/48\pi$. Therefore
\be\label{6.8}
3\tilde\kappa_{k=0}^2=3\tilde\kappa_\Lambda^2-1=25-c
\qquad (m_\Lambda=0)\ee
The combined theory $\Gamma_0^L
+\Gamma_0^{\rm grav}$ is conformal with total central charge zero.
By eq. (\ref{2.17}) the $\phi$-sector contributes $25-c$ and the
gravity sector $-(25-c)$. This has to be contrasted with
$m_\Lambda\not=0$ where the net central charge zero arises form
a compensation of $26-c$ in the $\phi$-sector and $-(26-c)$ in the
pure gravity sector. Here the ``pure gravity sector''
is a substitute for the integrated-out matter
and ghost system. In this sense we can confirm the usual statement
that the $\phi$-field contributes to the central charge like
an ordinary free boson, but we have also seen that depending
on whether $m_\Lambda=0$ or $m_\Lambda\not=0$ the mechanism
of how this comes about is quite different.

\section{The UV-Fixed Point}
\setcounter{equation}{0}

In this section we put Liouville field theory in a
more general context by looking at the space of 2-dimensional
theories with one scalar field as a whole.
In general, a critical scalar theory is governed by a fixed
point which typically is IR-unstable in one or several
directions. The flow starts then for $k=\Lambda$ in the
immediate vicinity of this fixed point and remains there
for a long ``running time''. Depending on the value of
$\phi$ the running stops at a certain point,
and $\Gamma_k[\phi]$ becomes
independent of $k$ for $k\to0$. Generically there is also
a region in $\phi$ where the running never stops. In the
critical ${\bf Z}_2$-symmetric $\phi^4$-theory
(Ising model) this region shrinks to one point --
the potential minimum -- as $k\to0$. In our case this region
is $\phi<\phi_c(k)$ and it moves to $\phi\to-\infty$ as
$k\to0$. We have argued in sect. 5 that the $k$-independent
form of $\Gamma_{k\to0}[\phi]$ is essentially the classical
Liouville action. In this section we want to investigate
closer the fixed point which governs the scaling behavior.
In particular,
we search for fixed points of the renormalization group
flow for appropriate dimensionless variables.
Though the fixed point equation is easier to solve
than the evolution equation, it is impossible to find
exact solutions unless one introduces approximations of some
sort. In the following we truncate the space of actions
by considering only functionals of the form
(``local potential approximation'')
\be\label{7.1}
\Gamma_k[\phi;g]=Z_k\int d^2x\sqrt g\left\{\frac{1}{2}
D_\mu\phi D^\mu\phi+Rf_k(\phi)+k^2u_k(\phi)\right\}\ee
Here $f_k$ and $u_k$ are arbitrary dimensionless functions.
Their evolution is governed by (\ref{4.1}),
\bear\label{7.2}
&&Z_k\int d^2x\sqrt g\Bigl\{\partial_tu_k+2u_k+\partial_tf_k
\ k^{-2}\ R\nonumber\\
&&\nonumber\\
&&+Z_k^{-1}\partial_tZ_k[\frac{1}{2}k^{-2}(D_\mu\phi)^2+k^{-2}
\ R\ f_k+u_k]\Bigr\}\nonumber\\
&&\nonumber\\
&&=k^{-2}{\Tr}\left[\left\{(1-\frac{1}{2}\eta)C+\frac{D^2}
{k^2}C'\right\}\left[-\frac{D^2}{k^2}+C+\frac{R}
{k^2}f_k''+u_k''\right]^{-1}\right]
\ear
with $C\equiv C(-D^2/k^2)$, and $f_k'',u_k''$
denotes the second derivatives with respect to $\phi$. The
interesting question is whether there exist fixed points
$(u_*,f_*,Z_*)$ in the subspace of actions
with the form (\ref{7.1}). Because all parameters are chosen
dimensionless, the condition is simply
$\partial_t u_k=\partial_t f_k=\partial_tZ_k=0$.
We emphasize that this is not the most general form of the fixed
point even within the space of actions given by our truncation
(\ref{7.1}). In fact, for general scalar theories
the anomalous dimension $\eta$ does not vanish at the fixed
point. In this case one should replace $\phi$ by $k$-dependent
renormalized fields $\phi_R=Z^{1/2}_k\phi$ in
order to obtain the appropriate scaling solution. The
condition $\partial_tZ_k=0$ selects therefore only the subclass
of fixed point solutions with vanishing $\eta$. In view of
the results of sects. 4 and 5 this is appropriate for
``Liouville-type'' critical scalar models.

For the determination of the fixed point potential $u_*(\phi)$
we may set $g_{\mu\nu}=\delta_{\mu\nu}$ and evaluate
the trace in momentum space:
\be\label{7.3}
8\pi Z_*\ u_*(\phi)=\int^\infty_0dy\frac
{C(y)-yC'(y)}{y+C(y)+u_*''(\phi)}\ee
Note that by introducing $y\equiv p^2/k^2$ the equation has
become independent of $k$, as it should. The r.h.s. of (\ref{7.3})
is not independent of $C$. For an exponential cutoff,
say, the $y$-integral converges as it stands. For simplicity
we use here $C\equiv1$ and subtract from
$u$ a constant which is not relevant  for our
purpose. With $\bar u(\phi)=u_*(\phi)-u_*(\phi_0)$
(where $\phi_0$ is arbitrary provided $u''_*(\phi_0)
>-1)$ eq. (\ref{7.3}) can
be written as
\be\label{7.4}
\frac{d^2}{d\phi^2}\bar u(\phi)=-\frac{d{\cal P}}{d\bar u}
(\bar u(\phi))\ee
with
\be\label{7.5}
{\cal P}(\bar u)\equiv \bar u+p_0e^{-8\pi Z_*\bar u}\ee
where $p_0\equiv(1+u_*''(\phi_0))/(8\pi Z_*)$.
With the identification
$\bar u(\phi)\hat= x(t)$ and ${\cal P}(\bar u)
\hat=V(x)$, eq. (\ref{7.4}) is Newton's equation for a particle
located at $x(t)$ and moving in the potential $V(x)$. Because
${\cal P}(\bar u)$ has a single minimum and ${\cal P}
(\bar u\to\pm\infty)\to+\infty$, all
solutions of (\ref{7.4}) with an ``energy''
\be\label{7.6}
{\cal E}\equiv\frac{1}{2}\left(\frac{d\bar u}{d\phi}\right)^2
+{\cal P}(\bar u)\ee
larger than the minimum are oscillatory, i.e. $\bar u(\phi)$
is a periodic function\footnote{This is also true
for an arbitrary cutoff function $C$.} of $\phi$. Because the fixed
points $\Gamma_*$ correspond to conformal field theories, one
might have hoped to recover Liouville theory in
this manner, but clearly an exponential $\bar u(\phi)\sim\exp(2\phi)$ is
not a solution of (\ref{7.4}). In fact, the oscillatory
solutions of (\ref{7.4}) were discussed in ref. \cite{mor}
already, where they were identified with the critical Sine-Gordon
models. The only solution which is not oscillatory is the constant one,
$\bar u(\phi)=\bar u_{\rm min}$, where the ``particle'' sits for
alle ``times'' $\phi$ at the minimum $\bar u_
{\rm min}$ of ${\cal P}$. It
corresponds to a Gaussian (or more appropriately, Feigin-Fuks)
fixed point for which the potential vanishes up to a constant.
We are now going to show that Liouville theory can be
understood as the perturbation of this fixed point by a relevant
operator. When $k$ is lowered from infinity down to zero, this
operator drives the theory from the IR unstable Gaussian fixed
point to a different IR stable one
which represents {\it quantum} Liouville theory and corresponds
to our solution for $k=0$.

For an arbitrary cutoff function $C$ the
constant fixed point solution reads
\be\label{7.7}
u_*=\frac{1}{8\pi Z_*}\int^\infty_0dy\frac{C(y)-yC'(y)}{y+C(y)}\ee
It singles out a point in the space of all actions, and we are
interested in the linearized flow in the tangent space at this
very point. We write
\be\label{7.8}
u_k(\phi)=u_*+\varepsilon\delta u_k(\phi)\ee
with $(t\equiv\ln (k/\Lambda))$
\be\label{7.9}
\delta u_k(\phi)=e^{-\gamma t}\ \Upsilon(\phi)+\sum_i
\ e^{-\gamma_it}\ e^{2\Delta^0_it}
\ \psi_i\  \Upsilon_i(\phi)\ee
and expand the evolution equation to first order in $\varepsilon$.
Here $\Upsilon(\phi)$ and $\Upsilon_i(\phi)$ are the eigenvectors of the
linearized evolution operator with the eigenvalues
$\gamma$ and $\gamma_i$, respectively. They are
obtained explicitly by inserting (\ref{7.8}) with (\ref{7.9})
into (\ref{7.2}) with $g_{\mu\nu}=\delta_{\mu\nu}$. The
terms independent of $\varepsilon$ cancel by virtue of (\ref{7.7}),
and the terms linear in $\varepsilon$ involve a momentum space
integral which is precisely of the type (\ref{4.9}). It is
independent of the form of $C$. Thus one finds
\be\label{7.10}
\Upsilon''(\phi)+4\pi Z_*(2-\gamma)\Upsilon(\phi)=0\ee
\be\label{7.11}
\Upsilon''_i(\phi)+4\pi Z_*(2-\gamma_i+2\Delta_i^0)\Upsilon_i(\phi)=0\ee
At this point we understand what is so special about the exponential
interaction potentials: They are the eigenvectors of the linearized
renormalization group flow at the Gaussian fixed point. Parametrizing
the solutions as $\Upsilon\sim \exp(2\alpha_*\kappa_*\phi)$ and
$\Upsilon_i\sim\exp(2\alpha_{i*}\kappa_*\phi)$,
and using $4\pi Z_*=\kappa^2_*$ we obtain
\be\label{7.12}
\gamma=2+4\alpha^2_*\ee
\be\label{7.13}
\gamma_i=2+4\alpha^2_{i*}+2\Delta^0_i\ee
Hence, near the fixed point, the potential scales according to
\be\label{7.14}
U_k(\phi)=
\varepsilon Z_*k^2\delta u_k(\phi)=\frac{m_\Lambda^2}
{16\pi\alpha_*^2}\left[\left(\frac
{\Lambda}{k}\right)^{4\alpha^2_*}e^{2\alpha_*\kappa_*\phi}+
\sum_i\psi_i\left(\frac
{\Lambda}{k}\right)^{4\alpha^2_{i*}}e^{2\alpha_{i*}\kappa_*\phi}
\right]\ee
where we have adapted the free multiplicative integration
constant to the notation (4.16).
Obviously, for $\alpha_*,\alpha_{i*}$ real,
the perturbation grows for
decreasing $k$ and the fixed point is IR-unstable. If we
identify $\alpha_*,\alpha_{i*}$ and $\kappa_*^2$ with the
UV values $\alpha_\Lambda,\alpha_{i\Lambda}$ and $\kappa_\Lambda$,
then eq. (\ref{7.14}) coincides precisely with the solution
of the evolution equation in section 5. This shows that the
truncation employed there is rather accurate for large
values of $k$ when the flow is still well
approximated by its linearization around the Gaussian
fixed point.

We have not determined the function $f_*(\phi)$ yet. Because
at the fixed point the l.h.s. of (\ref{7.2}) contains no
$\sqrt gRf_k(\phi)$ term, the trace on the r.h.s. must not
generate such a term. A sufficient condition for this not
to happen is that $f_*(\phi)=c_1+c_2\phi$. Imposing
in addition that the fixed point action $\Gamma_*$ satisfies
the Ward identity we see that $\Gamma_*$ is precisely the
Feigin-Fuks-type theory (\ref{6.5}). The coefficient
$\kappa_*^2$ is not determined by the fixed point condition,
but the Ward identity requires it to be
$3\kappa_*^2=3\kappa_\Lambda^2
=25-c+\tau$.

To conclude, let us summarize the general picture as it arises form
our investigation of the Ward identities, the evolution equation
and the fixed point equation. The renormalization group trajectory
starts at $k=\Lambda$ at a conformal theory with central charge
$3\kappa^2_\Lambda=25-c$ (for the Weyl invariant measure). This
initial point approaches for $\Lambda\to\infty$
a Feigin-Fuks free field action
$\Gamma_\Lambda^L(m=0)$ as in eq. (\ref{6.5}). The relevant
perturbations at this fixed point are governed by eigenvectors
of the linearized flow, $\Upsilon(\phi)$ and
$\Upsilon_i(\phi)$, which have an
exponential dependence on $\phi$. Adding such a perturbation
to $\Gamma_\Lambda^L(m=0)$ we obtain the Liouville action. The
perturbation drives the system away from the IR-unstable fixed
point. For $k\to0$, $\Gamma_k^L$ approaches an IR-stable
fixed point, a conformal theory with central
charge $26-c$.\footnote{This is not in
contradiction with the Zamolodchikov
$c$-theorem because $k$ is an IR cutoff.} The exact Weyl Ward identity
guarantees that $\Gamma_{k\to0}^L$ is conformal
with $c[\Gamma_{k\to0}]=26-c$. With our truncation we find
that $\Gamma_0^L$ equals the classical Liouville action.
We expect that this is a very good approximation (and
perhaps even exact) for the low momentum part of the effective
action.

In this context it is important to keep in mind that the area
operator is required to be an $(1,1)$ operator in the quantum
theory, i.e., for $k\to0$, but not at $k=\Lambda$. In the
UV it is not marginal and this is what drives the system away from
the UV-fixed point. In fact, to a conformal reparametrization
$z\to z'(z)$, $d{s'}^2=|dz'/dz|^2ds^2,\phi'=\phi-\frac{1}{2}\ln
|dz'/dz|^2$ the area operator responds according to
\be\label{7.15}
(\sqrt ge^{2\alpha_\Lambda\kappa_\Lambda\phi})'
=\left|\frac{dz'}{dz}\right|^{-4\alpha^2_\Lambda}\sqrt ge^{2\alpha
_\Lambda\kappa_\Lambda\phi}\ee
with the familiar exponent $4\alpha^2_\Lambda$. This is also
reflected by the presence of the field-dependent terms in
$T^\mu_\mu[\Gamma_\Lambda^L]$ which we found in section 3.

It is interesting that recently evidence for a similar mechanism
in $d>2$ was found \cite{hh}, \cite{per}. Halpern
and Huang investigated the linearized renormalization group flow
at the Gaussian fixed point in O(N) symmetric scalar theories.
They find non-polynomial interactions for which the Gaussian fixed
point is IR-unstable rather than IR attractive (as in the
case of polynomial potentials). Hence these theories are
asymptotically free rather than trivial. It is tempting to
regard them as higher dimensional counterparts of Liouville
theory. In fact, for $d>2$ the potentials behave as $U(\phi)
\sim\exp[a(d-2)\phi^2]$ for large $\phi$, where $a$ is a
constant. These results have been questioned in ref. \cite{trm},
however.

\section{Conclusion}

In this paper we have developed the general framework for the
quantization of Liouville theory using exact renormalization
group equations. We applied this program to a specific truncation
of the space of actions. An important role was played by the
Ward identities resulting from the Weyl invariance of the
theory. They imply that the exact renormalization group trajectory
$\Gamma_k$ runs in the IR into a fixed point which is a conformal
field theory of central charge $26-c$. For $k\to\infty$ the Ward
identities give rise to constraints on the initial values of the
parameters contained in $\Gamma_\infty$. Only if these constraints
are satisfied, the evolution drives $\Gamma_k$ towards the
correct IR fixed point. Other initial values would correspond
to other universality classes. In particular, the initial
values for $\alpha_\Lambda$ and $\alpha_{i\Lambda}$ are
fixed in terms of $c$ and the bare scaling dimensions
$\Delta^0_i$ by eqs. (3.31) and (\ref{3.28}). These formulas
are well known from the standard approaches where they arise
from the
requirement that the area operator and the gravitationally dressed
matter operator have dimensions $(1,1)$ in the {\it quantum} theory
so that they can be integrated over the Riemann surface. In our
case the interpretation is different: A priori
these relations are statements about the microscopic or
{\it classical}
theories $\Gamma_\infty$ which are within the ``range of attraction''
of the correct IR fixed point, i.e., the evolution (quantization)
is still to be performed. In section 4 we obtained the solution
of the truncated evolution equation, and we saw that the
gravitationally modified dimensions $\Delta_i(k)$ are
renormalization group invariants only for large enough $k$.
Only in this range
the KPZ formula is reproduced. For $k\to0$
the dimensions $\Delta_i(k)$ shift back to the
``bare'' values $\Delta_i^0$, which apply without quantized gravity.

As for our solution of the evolution equation, it is quite impressive
to see the high degree of ``conspiracy'' among the $\beta$ functions
of $\alpha_k,\kappa_k,\tilde\kappa_k$ and $m_k$. When
$k$ is lowered, $m_k$ grows precisely at a rate which
cuts off the running of $\kappa_k$ in such a manner that
it changes during an infinitely long running time by a finite
amount only. Though the Liouville ``mass'' is crucial for this
mechanism, the same shift obtains also
for $m_\Lambda=0$. The $\beta$-functions of $\kappa_k$
and $\tilde\kappa_k$  conspire in such a way that for
$m_\Lambda=0$ the task of changing the central charge
can be transferred from
$\kappa_k$ to $\tilde\kappa_k$.

In this paper we used a truncation which is inspired by the classical
Liouville action. It contains the same types of operators but
with $k$-dependent coefficients. In the case of standard
(scalar) theories truncations of this type were already
successfully used in a variety of applications \cite{ng},
\cite{mor}, \cite{gau}. We have checked the quality of the
truncation by comparing the solution of the evolution equation
to the constraints imposed by the Ward identities. It turned out
to be very satisfactory both for $k\to\infty$ and $k\to0$.
We believe that our truncation indeed reflects the most
important qualitative and quantitative features
of the quantum Liouville theory in the range considered,
i.e., for $c\leq 1$ and for low momenta.

By combining the informations about the renormalization-group
trajectory obtained from the Ward identities, the evolution
equation and the fixed-point equation we are led to consider
Liouville field theory for $c<1$ as a crossover phenomenon
from one renormalization group fixed point to another. The
trajectory starts at an UV-stable Feigin-Fuks (``Gaussian'')
fixed point. The area operator is a relevant operator there
and it drives the trajectory towards a second, IR-stable
fixed point, the quantum Liouville theory with central charge
$26-c$. In our truncation the effective action of the latter
equals the classical Liouville action with a
renormalized parameter $m^2$.

It remains an important open question how to improve the
truncation for $k\to0$, and how to recover the KPZ scaling laws,
for instance.
It is quite clear that finding such an
improved truncation will be very difficult because calculating
arbitrary correlators in Liouville theory is known to be
a formidable task, and these correlators contain the same
information as $\Gamma_{k\to0}$. If one naively adds further
terms to the ansatz for $\Gamma_k$ one sees immediately that it
is very hard to produce to ``good'' truncation
which does not lead to inconsistencies with the Ward identities.

It should be emphasized that the present work deals essentially
with the low momentum behavior of quantum Liouville theory.
This is manifest in the study of the effective potential
which generates the $n$-point functions at zero momentum.
Also the investigation of the kinetic term was restricted
to the lowest order in a derivative expansion and therefore
to the limit of the lowest order of a Taylor expansion
around vanishing momentum. It is a very interesting question
what happens to $n$-point functions with typical external
(euclidean) momentum squared $q^2>0$. The main
observation here is that, for generic external momenta, $q^2$ acts
as an effective infrared cutoff, very similar to $k^2$. We
therefore suggest that for $q^2>0$ the $n$-point functions
of quantum Liouville theory $(k=0)$ are qualitatively described
by the generating functional $\Gamma_k$ evaluated at zero
momentum, and with $k^2$ chosen such that $k^2\approx q^2$.
In other words, the role of the effective infrared cutoffs
$k^2$ and $q^2$ can be interchanged. If this simple picture
holds true, our study of the effective action for
arbitrary $k$ provides important information on the
momentum dependence of the $n$-point functions of quantum
Liouville theory. In particular, we expect a different behavior
for $q^2$ larger or smaller than $q^2_c \equiv \bar m^2\exp(2\phi)$.
(Note the $\phi$-dependence of the critical momentum $q_c^2$,
which is necessary for a conformal theory.) For $q^2<q_c^2$
the $n$-point functions are essentially given by $\Gamma_{k\to0}$
as discussed in this work, i.e., the classical Liouville action.
On the other hand, large momenta $q^2>q_c^2$ should be described by
the $k\to\infty$ limit of $\Gamma_k$. It is in this region
of large momenta that the operators $\psi_i$ scale according
to the KPZ rule. This observation leads naturally to the idea of
effective momentum-dependent dimensions $\Delta_i(q^2)$, in close
analogy to $\Delta_i(k)$. Furthermore, one expects a shift in
the effective value of $\kappa$ and $\alpha\kappa$
as $q^2$ crosses the threshold $q^2_c$. Finally the factor
$k^{-4\alpha_\Lambda^2}$ in front of the exponential
potential should now be translated into an effective
momentum-dependent form factor which behaves for generic
momenta as $(q^2)^{-2\alpha_\Lambda^2}$.
The precise behavior near the threshold $q_c^2$ is difficult,
but in order to get an idea of the qualitative behavior we
can approximate the threshold effects by a form factor
$\left(\frac{q^2+q^2_c}{q^2_c}\right)^{-2\alpha_\Lambda^2}$.
As an example, this would lead to a characteristic behavior
of the inverse propagator as
\be\label{Z}
q^2+\Bigl[\bar m^2\exp(2\beta(q^2)\phi)\Bigr]^{1+2\alpha_\Lambda^2}
\ \Bigl[q^2+\bar m^2\exp(2\beta(q^2)\phi)\Bigr]^{-2\alpha^2_\Lambda}\ee
with $\beta(q^2\gg q_c^2)=\alpha_\Lambda\kappa_\Lambda$
and $\beta(q^2\ll q_c^2)=1$. In particular, this
propagator becomes a massless propagator for $q^2\to\infty$.
All this would have important implications for the general
picture of $n$-point functions in quantum Liouville theory
and certainly merits further study.

In most of the discussion we had to restrict ourselves to $c\leq1$.
In our approach the ``$c=1$ barrier'' is encountered if one
tries to find an admissible initial point $\Gamma_\infty$
within the truncated space of actions. As a consequence of
the Ward identity, $\alpha_\Lambda\kappa_\Lambda$ is
complex for $1<c<25$ and such a point simply does not
exist. Hence, if it exists at all, $\Gamma_\infty$ should be
of a form rather different from the Liouville action. Here we
are in the somewhat unusual situation that the quantization
of a theory has very little to do with its classical action:
$S_L$ enters only via its symmetry properties in the derivation
of the Ward identity. Of course, by the very idea of universality,
this is not a problem of principle but rather the generic
situation: Universality classes are fixed essentially by the field
content and the symmetries of a theory. The precise form
of the initial action does not matter as long as it is
attracted towards the correct IR fixed point. However, from a purely
practical point of view it is quite disturbing that the classical
action is of so little help in finding a representative of the
desired universality class. It is one of the advantages of our
formulation that the technically simpler step of finding $\Gamma_\infty$
is decoupled from the problem of the actual quantization (evolution).
An investigation of the region $1<c<25$ has to start with an
examination of the Ward identity (\ref{2.22}) for a truncation
as general as computationally feasible. The technically difficult
part is the evaluation of ${\A}_k,k\to\infty$,
for the assumed form of $\Gamma_k$, see eq. (\ref{2.24}). If
rather complicated nonlocal functions of $-D^2$ become
important, $\Gamma_k^{(2)}$ is a complicated nonminimal
differential operator in general, and almost nothing
is known about the heat-kernel expansion of $\Gamma_k^{(2)}$
in a fairly general form. Thus any progress in this direction
depends crucially on our ability to handle the Seeley-DeWitt expansion
of such nonminimal operators. (For $k\to\infty$ it is sufficient to
know the first few terms in this expansion in order to calculate
${\A}_k(x)$.)

In conclusion we think that the investigation
of exact evolution equations is a
promising new approach which puts Liouville theory into a novel
perspective and provides new calculational tools which have not
been exploited so far.

\bigskip
\noindent{\bf Acknowledgement:} M. R. would like to thank
G. Weigt for the numerous discussions which triggered these
investigations. He is also grateful to A. Bonanno,
M. Bonini, H. Dorn, H. Nicolai, J. Teschner, N. Tetradis
and A. Wipf for helpful conversations.
\newpage
\renewcommand{\baselinestretch}{1}

\end{document}